\RequirePackage{lineno}
\documentclass[twocolumn,letterpaper,aps,prc,longbibliography,superscriptaddress,nofootinbib,floatfix]{revtex4-1}

\usepackage{amsmath}	
\usepackage{graphicx}	
\usepackage{hyperref}
\hypersetup{colorlinks=true, urlcolor=cyan, linkcolor=blue, citecolor=blue}

\usepackage{xspace}	

\newcommand{\raa}{\mbox{$R_{AA}$}\xspace}
\newcommand{\rpa}{\mbox{$R_{pA}$}\xspace}

\newcommand{\Npart}{\mbox{$N_{\rm part}$}\xspace}

\newcommand{\snn}{\mbox{$\sqrt{s_{_\mathrm{NN}}}$}\xspace}
\newcommand{\sqsn}{\mbox{$\sqrt{s_{_\mathrm{NN}}}$}\xspace}

\newcommand{\pp}{\mbox{$p$+$p$}\xspace}
\newcommand{\pA}{\mbox{$p$+$A$}\xspace}
\newcommand{\pPb}{\mbox{$p$+Pb}\xspace}
\newcommand{\PbPb}{\mbox{Pb+Pb}\xspace}
\newcommand{\pO}{\mbox{$p$+O}\xspace}
\newcommand{\OO}{\mbox{O+O}\xspace}

\newcommand{\jpsi}{\mbox{$J/\psi$}\xspace}

\def\Ups#1{\mbox{$\Upsilon(#1 S)$}\xspace}
\def\pt{\mbox{$p_T$}\xspace}

\def\vtwo{\mbox{$v_2$}\xspace}
\def\Npart{\mbox{$N_{\rm part}$}\xspace}

\begin{document}

\title{Model study on $\Upsilon(nS)$ modification in small collision systems}

%
\newcommand{\inha}{Department of Physics, Inha University, Incheon 22212, South Korea}
\affiliation{\inha}

\newcommand{\yonsei}{Department of Physics and Institute of Physics and Applied Physics, Yonsei University, Seoul 03722, Korea}
\affiliation{\yonsei}

\newcommand{\jeonbuk}{Division of Science Education, Jeonbuk National University, Jeonju 54896, South Korea}
\affiliation{\jeonbuk}

\newcommand{\pusan}{Department of Physics, Pusan National University, Busan, 46241, South Korea}
\affiliation{\pusan}

\newcommand{\sejong}{Department of Physics, Sejong University, Seoul, 05006, South Korea}
\affiliation{\sejong}

\newcommand{\korea}{Department of Physics, Korea University, Seoul 02841, South Korea}
\affiliation{\korea}

\author{Junlee Kim} \affiliation{\jeonbuk}
\author{Jinjoo Seo} \affiliation{\inha}
\author{Byungsik Hong} \affiliation{\korea}
\author{Juhee Hong} \affiliation{\yonsei}
\author{Eun-Joo Kim} \affiliation{\jeonbuk}
\author{Yongsun Kim} \affiliation{\sejong}
\author{MinJung Kweon} \affiliation{\inha}
\author{Su Houng Lee} \affiliation{\yonsei}
\author{Sanghoon Lim} \affiliation{\pusan}
\author{Jaebeom Park} \affiliation{\korea}

\date{\today}

\begin{abstract}

Quarkonium production has been studied extensively in relativistic heavy-ion collision experiments to understand the properties of the quark gluon plasma.
The experimental results on the yield modification in heavy-ion collisions relative to that in $p$+$p$ collisions can be described by several models considering dissociation and regeneration effects.
A yield modification beyond initial-state effects has also been observed in small collision systems such as $p$+Au and $p$+Pb collisions, but it is still premature to claim any hot medium effect.
A model study in various small collision systems such as $p$+$p$, $p$+Pb, $p$+O, and O+O collisions will help quantitatively understanding nuclear effects on the $\Upsilon(nS)$ production. 
A theoretical calculation considering the gluo-dissociation and inelastic parton scattering and their inverse reaction reasonably describe the suppression of $\Upsilon(1S)$ in Pb+Pb collisions. 
Based on this calculation, a Monte-Carlo simulation is developed to more realistically incorporate the medium produced in heavy-ion collisions with event-by-event initial collision geometry and hydrodynamic evolution.
We extend this framework to small systems to study the medium effects.
In this work, we quantify the nuclear modification factor of $\Upsilon(nS)$ as a function of charged particle multiplicity ($dN_{ch}/d\eta$) and transverse momentum. We also calculate the elliptic flow of $\Upsilon(nS)$ in small collision systems.

\end{abstract}


\maketitle

\section{Introduction}
\label{sec:Intro}

Quarkonia have long been considered as golden probes to study the strongly interacting matter consisting of deconfined quarks and gluons, the quark-gluon plasma (QGP), produced in high-energy heavy-ion collisions~\cite{Busza:2018rrf,Karsch:2000ps,Shuryak:1977ut,Matsui:1986dk,Digal:2001ue}. 
Quarkonium states are produced at the early stages of the collision via hard parton scatterings, thus experiencing the full space-time evolution of the medium.
Also, their spectral functions are modified due to color screening~\cite{Matsui:1986dk,Digal:2001ue} and interactions with medium constituents such as gluo-dissociation or Landau damping~\cite{laine:2007,Brambilla:2008cx,Brambilla:2010vq}.
Consequently, the quarkonium yields are expected to be suppressed in heavy ion collisions with respect to expectations from proton-proton (\pp) data, following the order of their binding energies.
On the other hand, the yields of quarkonia can be enhanced in the presence of the QGP by recombination processes of uncorrelated as well as correlated quarks~\cite{Gorenstein:2000ck,Andronic:2007bi,Ravagli:2007xx,blaizot:2016jp}.

The modification of the quarkonium yields have been studied by various experiments at RHIC and LHC using the nuclear modification factor quantified as the yield ratio in nucleus-nucleus collisions ($A$+$A$) to that in \pp collisions scaled by the average number of binary $NN$ collisions~\cite{STAR:2013kwk,CMS:2017uuv,ATLAS:2018hqe,CMS:2018zza,ALICE:JpsiRAA2019,ALICE:2021UpsForward,ATLAS:2022xso,CMS-PAS-HIN-21-007}. 
One of the most remarkable signatures is the ordered suppression of \Ups 1, \Ups 2, and \Ups 3 mesons by their binding energies reported in LHC~\cite{ALICE:2021UpsForward,ATLAS:2022xso,CMS:2018zza,CMS-PAS-HIN-21-007}.

To better understand the in-medium effects of quarkonia in $A$+$A$ collisions in a sophisticated way, it is important to study the ``cold nuclear matter" (CNM) effects which are typically probed using proton-nucleus (\pA) collisions. Modification of parton distribution functions in the nucleus~\cite{Vogt:2015uba}, energy loss~\cite{Arleo:2015lja} or nucleus absorption~\cite{McGlinchey:2012bp,Arleo:1999af}, and interactions with comoving particles~\cite{Capella:1996va,Ferreiro:2014bia,Ferreiro:2018wbd} are examples of CNM effects. 
On the other hand, various experiments have reported capital results, suggesting a QGP-like behavior of the created medium also in smaller collision systems, such as the observation of long-range collective azimuthal correlations in high multiplicity regions~\cite{CMS:2010ifv,ATLAS:2015hzw,CMS:2016fnw,ALICE:2021nir,CMS:2012qk,ALICE:2012eyl,LHCb:2015coe,CMS:2015yux,CMS:2019wiy,STAR:2015kak,PHENIX:2018lia}. 
Therefore, sophisticated phenomenological studies in such interactions become the subject that is sensitive to understanding the quarkonium production in small collision systems.

In this paper, we report a detailed study of the in-medium effects for \Ups 1, \Ups 2, and \Ups 3 mesons in proton-lead (\pPb), proton-oxygen (\pO), and oxygen-oxygen (\OO) collisions. 
Theoretical calculations for dissociation of \Ups n~\cite{Hong:2019ade} are incorporated with the SONIC framework~\cite{Romatschke:2015gxa} to describe the time evolution of the medium. 
The dissociation component is constraint in potential non-relativistic QCD (pNRQCD) limits, and coupled into the Boltzmann equation. The thermal width is calculated based on hard thermal loop (HTL) perturbation theory using the Bethe-Salpeter amplitude.
We report the nuclear modification factors and the second-order Fourier coefficient (\vtwo) of the azimuthal distribution of \Ups n mesons in \pPb, \pO, and \OO collisions, and the contribution of feed-down from higher excited states are considered to compare with the experimental data properly.
For the demonstration of the framework, we also present the results in \PbPb collisions and compare them with the experimental results.


\section{Simulation Framework}  
\label{sec:Simul}

The simulation framework is composed of two parts, hydrodynamics simulation for background medium and medium response of quarkonia.
For the first part, we follow the procedure described in Ref.~\cite{Lim:2018huo}.
Initial energy density in the transverse plane of in the collision of nucleus-nucleus or proton-nucleus is obtained with the MC-Glauber framework~\cite{Miller:2007ri}.
In the MC-Glauber, a nucleon-nucleon inelastic cross section of 72 mb~\cite{ATLAS:2016tc} is employed for small system collisions at \sqsn= 8 TeV at the LHC. 
A Gaussian of width $\sigma=0.4$ fm is used to describe the energy deposition of each nucleon participating in at least one inelastic collision.
The deposited energy distribution from all wounded nucleons in each event is converted into energy density for the hydrodynamic simulation (SONIC). 
A single scale factor is used for all events in a certain collision system, and the scale factor is determined to match the charged particle multiplicity ($dN_{ch}/d\eta$) at mid-rapidity in \pPb collisions at \snn = 8.16 TeV~\cite{CMS:2017shj,ALICE:2018wma}.
The same scale factor is used for \pO and \OO collisions at \snn = 8 TeV by assuming the scale factor does not change much in the collision systems with a similar number of participants.

\begin{figure}[htb]
    \centering
        \includegraphics[width=0.9\linewidth]{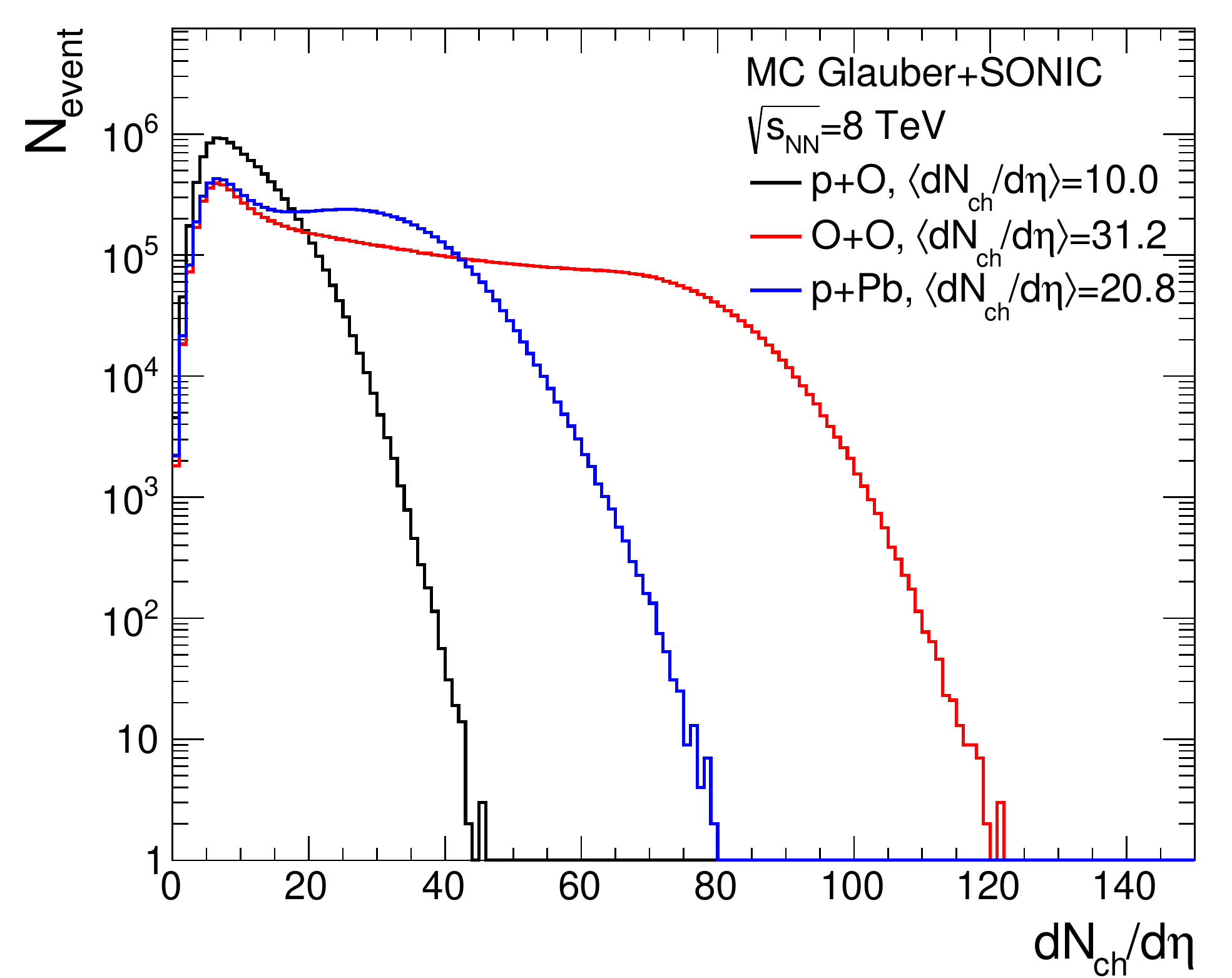}
    \caption{Charged particle multiplicity distribution in \pO, \OO, and \pPb collisions at $\sqsn=8$ TeV from SONIC.}
    \label{fig:sonic_mult_small}
\end{figure}

\begin{figure}[htb]
    \centering
        \includegraphics[width=0.9\linewidth]{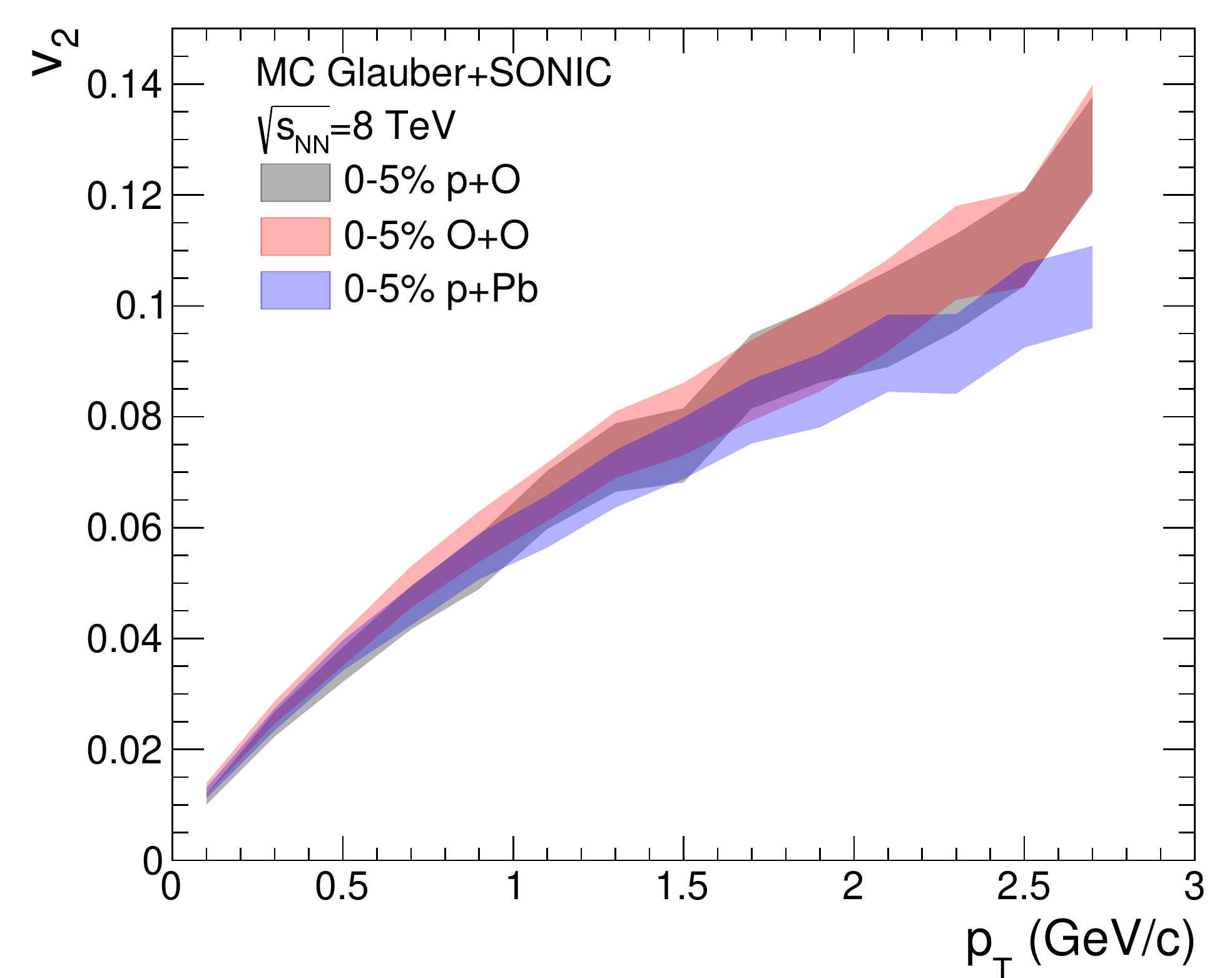}
    \caption{Elliptic flow as a function of \pt for charged particles in 0--5\% central \pO, \OO, and \pPb collisions at $\sqsn=8$ TeV from SONIC.}
    \label{fig:sonic_v2_small}
\end{figure}

Figure~\ref{fig:sonic_mult_small} shows the charged particle multiplicity distributions of unbiased (0--100\%) \pO, \OO, and \pPb collisions at \snn = 8 TeV from SONIC. 
The mean charged particle multiplicity for \pO and \OO with the scale factor obtained from \pPb collisions is 10.0 and 31.2, respectively.
We also check the elliptic flow of charged particle in 0--5\% high multiplicity events as shown in Fig.~\ref{fig:sonic_v2_small}, and the \pPb result is slightly lower than \pO and \OO results at higher \pt. 
Note that we use a shear viscosity to entropy density ratio of 0.08 and a bulk viscosity ratio of zero in the SONIC calculations.
There are other frameworks for initial conditions~\cite{Moreland:2014oya,Schenke:2012wb}, but we use only the MC-Glauber initial condition as a simple case in this study.

\begin{figure}[htb]
    \centering
        \includegraphics[width=0.8\linewidth]{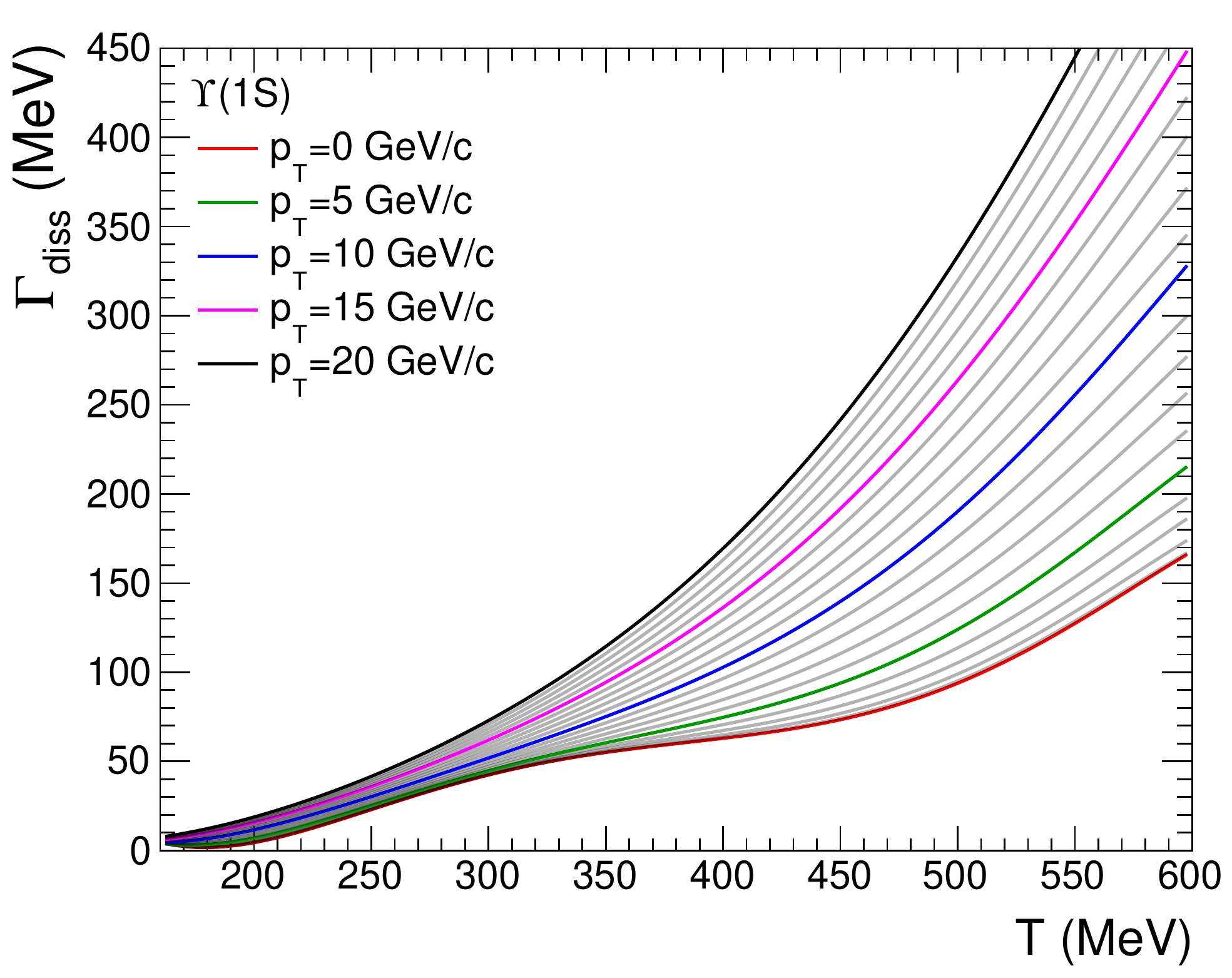}
        \includegraphics[width=0.8\linewidth]{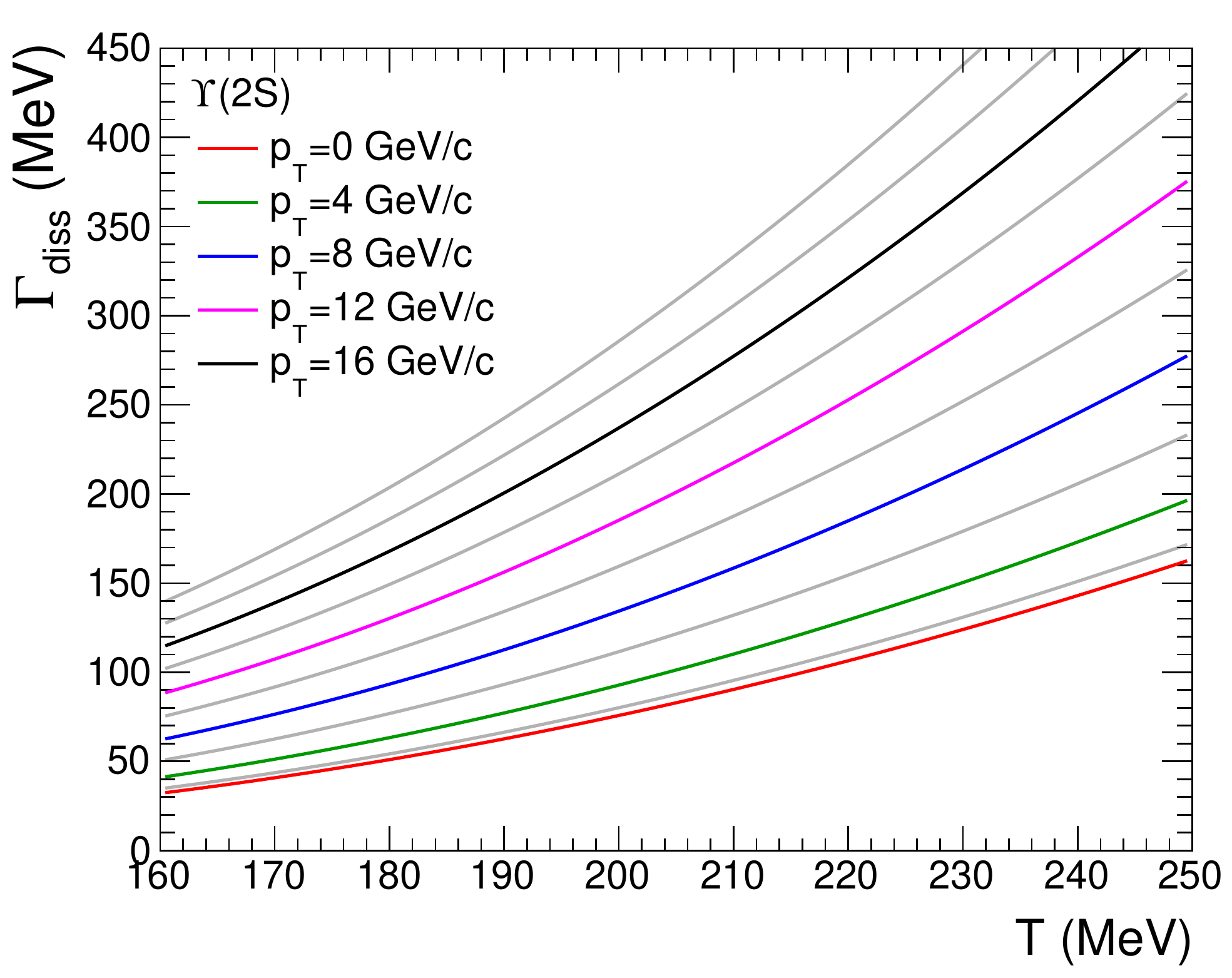}
        \includegraphics[width=0.8\linewidth]{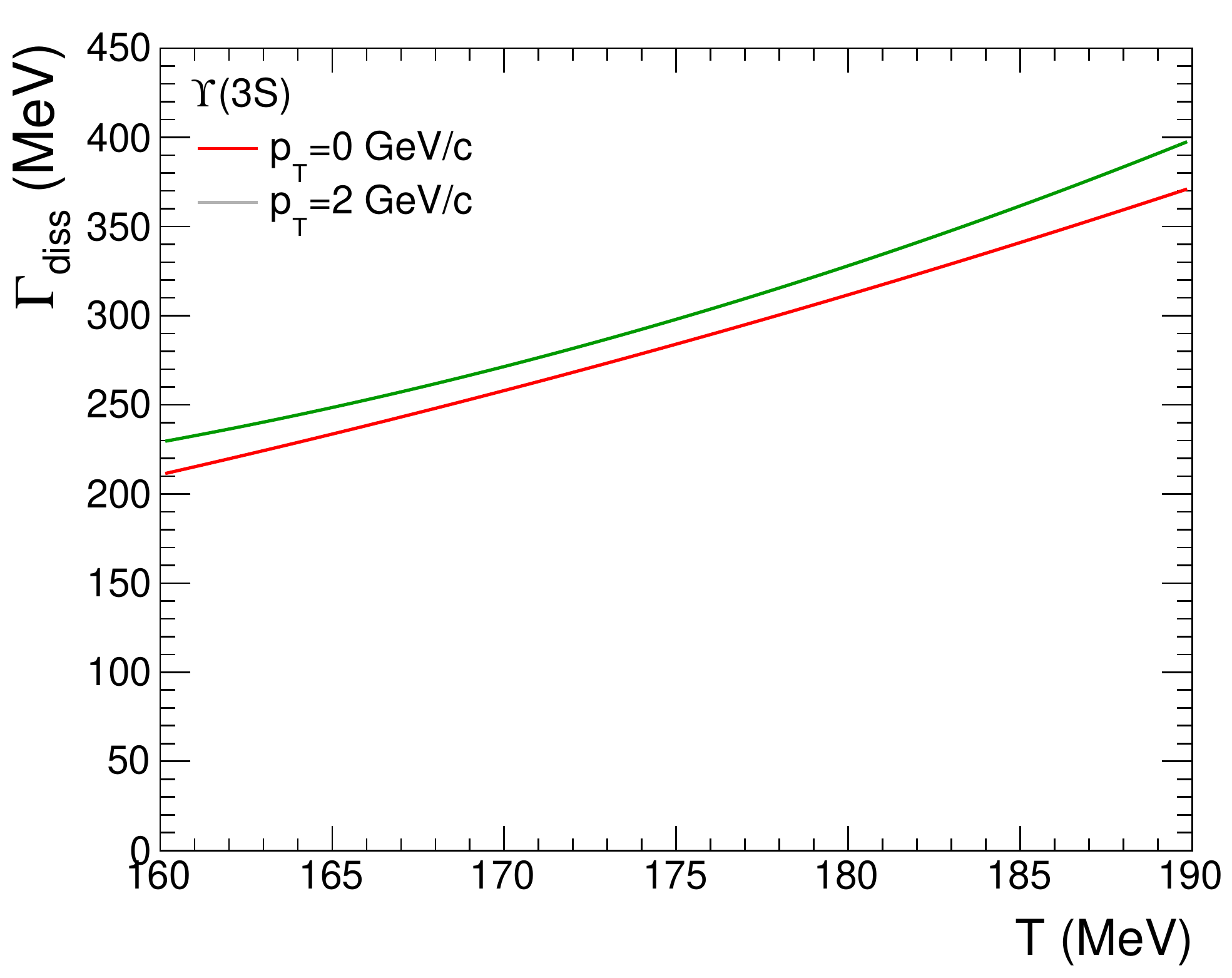}
    \caption{Thermal width as a function of temperature for different \pt of \Ups 1 (top), \Ups 2 (middle), and \Ups 3 (bottom).}
    \label{fig:thermal_width}
\end{figure}

During the SONIC simulation, temperature distributions at multiple time steps are stored for medium response of Upsilons.
At the beginning of the second part, we generate \Ups n mesons, and the $x$ and $y$ positions are determined based on the initial energy density distribution.
We used a Tsallis fit to the \pt distribution of \Ups 1 in $\sqrt{s}=5.02$ TeV~\cite{CMS:2018zza} to sample \pt of Upsilons, and $p_{x}$ and $p_{y}$ are determined based on a randomly assigned azimuthal angle. The initial \pt distribution is assumed to be the same for 5.02 TeV and 8 TeV. Also, any difference in the \pt shape, if any, is expected to be negligible because the results are obtained in fine \pt bins. 
The medium response is simulated based on the procedure described in Ref.~\cite{Hong:2019ade}.
The fraction of survived Upsilons for a certain time step ($\Delta t$) is calculated as:

\begin{align}
\label{eq:rate}
    \frac{N(t+\Delta t,p_{T})}{N(t,p_{T})} = e^{- \int^{t+\Delta t}_{t} d t^{\prime} \Gamma_\mathrm{diss}(t^{\prime},p_{T})},
\end{align}
where $\Gamma_\mathrm{diss}$ is the thermal width depending on the medium temperature and Upsilon \pt, and $\Delta t$ is set to 0.02 fm/$c$ in the hydrodynamic simulation.
The medium temperature is from the SONIC simulation according to $x$, $y$, and $t$ values. The temperature in which the quarkonium's in-medium binding energy reaches zero, i.e. dissociation temperature, is set as 600 MeV, 240 MeV, and 190 MeV for \Ups 1, \Ups 2, and \Ups 3, respectively~\cite{MPS:2013qqgp,HS:2006cdqb}.  
The thermal width from numerical calculations considering gluo-dissociation and inelastic parton scattering is obtained from Ref.~\cite{Hong:2019ade}.
Figure~\ref{fig:thermal_width} shows thermal width as function of temperature for \Ups 1 (left), \Ups 2 (middle), and \Ups 3 (right), and each line represents a distribution for different \pt.
It generally increases with \pt, and the thermal width for excite states is larger.
Note that thermal width for \Ups 3 is obtained at the limited \pt range, so we estimated values for $\pt>2~\mathrm{GeV}/c$ using the \pt dependence of \Ups 2 as:

\begin{align}
    \Gamma_\mathrm{diss}^{\Upsilon(3S)}&(p_{T}) = \Gamma_\mathrm{diss}^{\Upsilon(3S)}(2~\mathrm{GeV}/c) \frac{\Gamma_\mathrm{diss}^{\Upsilon(2S)}(p_{T})}{\Gamma_\mathrm{diss}^{\Upsilon(2S)}(2~\mathrm{GeV}/c)}.
\end{align}

A different formation time ($\tau_\mathrm{form}$) of each states is used as 0.5, 1.0, and 1.5 for \Ups 1, \Ups 2, and \Ups 3, respectively, based on values in Ref.~\cite{Du:2017qkv}.
We do not consider dissociation at the pre-resonance stage, so the medium response on the $\Upsilon$ states is turned off for $\tau< \gamma  \tau_\mathrm{form}$, where $\gamma$ is the Lorentz factor.
We found that the overall medium response is sensitive to the choice of the formation time and pre-resonance response. 
The setting that can well reproduce heavy-ion results is used for small systems consistently.
After incorporating the regeneration effect in heavy-ion collisions, a detailed systematic study will be performed later.

In each time step after the formation time, a survival rate is calculated for each $\Upsilon$ based on Eq.~\ref{eq:rate}, and the $\Upsilon$ is removed when a random number from the uniform distribution within 0--1 is greater than the survival rate. 
If the $\Upsilon$ is survived, the position for the next time step is calculated based on the momentum and time step ($\Delta t$).
We repeat the survival rate calculation with a different temperature at the new position until the temperature is lower than a critical temperature of 170 MeV.
For the simulation results presented in later sections, we use 1000 events for MC-Glauber and SONIC.
The number of generated \Ups n for a certain event scales with the number of inelastic collisions from the MC-Glauber. 
In this study, we do not consider a contribution from the regeneration effect which is not expected to be significant in small collision systems. 
In addition, there could be other nuclear effects such as modification of parton density, initial-state energy loss, and other final-state effects, but we focus on evaluating the dissociation effect.

\subsection{Nuclear modification factor}
The nuclear modification factors (\rpa and \raa) are defined as the ratios of the production cross sections of $\Upsilon$ mesons in \pA and $A$+$A$ collisions to the expected cross sections extrapolated from \pp collisions. To make an equivalent comparison, we took the ratio of generated to survived number of $\Upsilon$s in the full medium response simulation as the nuclear modification factor.

\subsection{Elliptic flow}
The elliptic flow ($v_2$) is calculated from the azimuthal angle distribution of survived \Ups n with respect to the event plane angle

\begin{align*}
    \frac{dN}{d(\phi - \Psi)} \propto 1 + 2v_{2}(p_T) \cos (2 (\phi - \Psi) ),
\end{align*}
where $\phi$ is the azimuthal angle of the \Ups n, and $\Psi$ is the event plane angle.
The event plane angle for each event is calculated with the initial energy density profile from the MC-Glauber.

\subsection{Feed-Down correction}
\label{sec:FDCorr}
\begin{figure}[htb]
    \centering
        \includegraphics[width=0.8\linewidth]{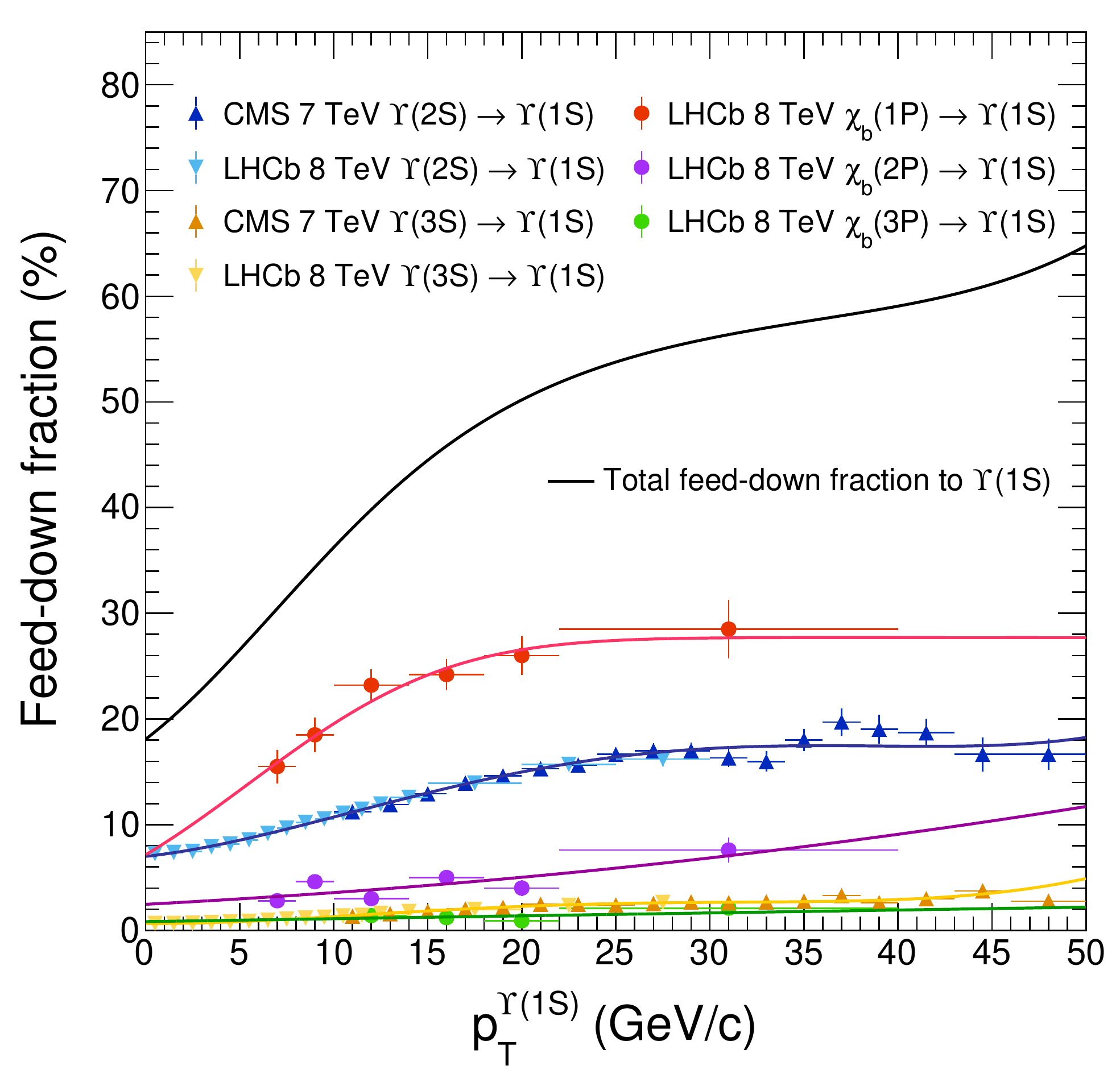}
        \includegraphics[width=0.8\linewidth]{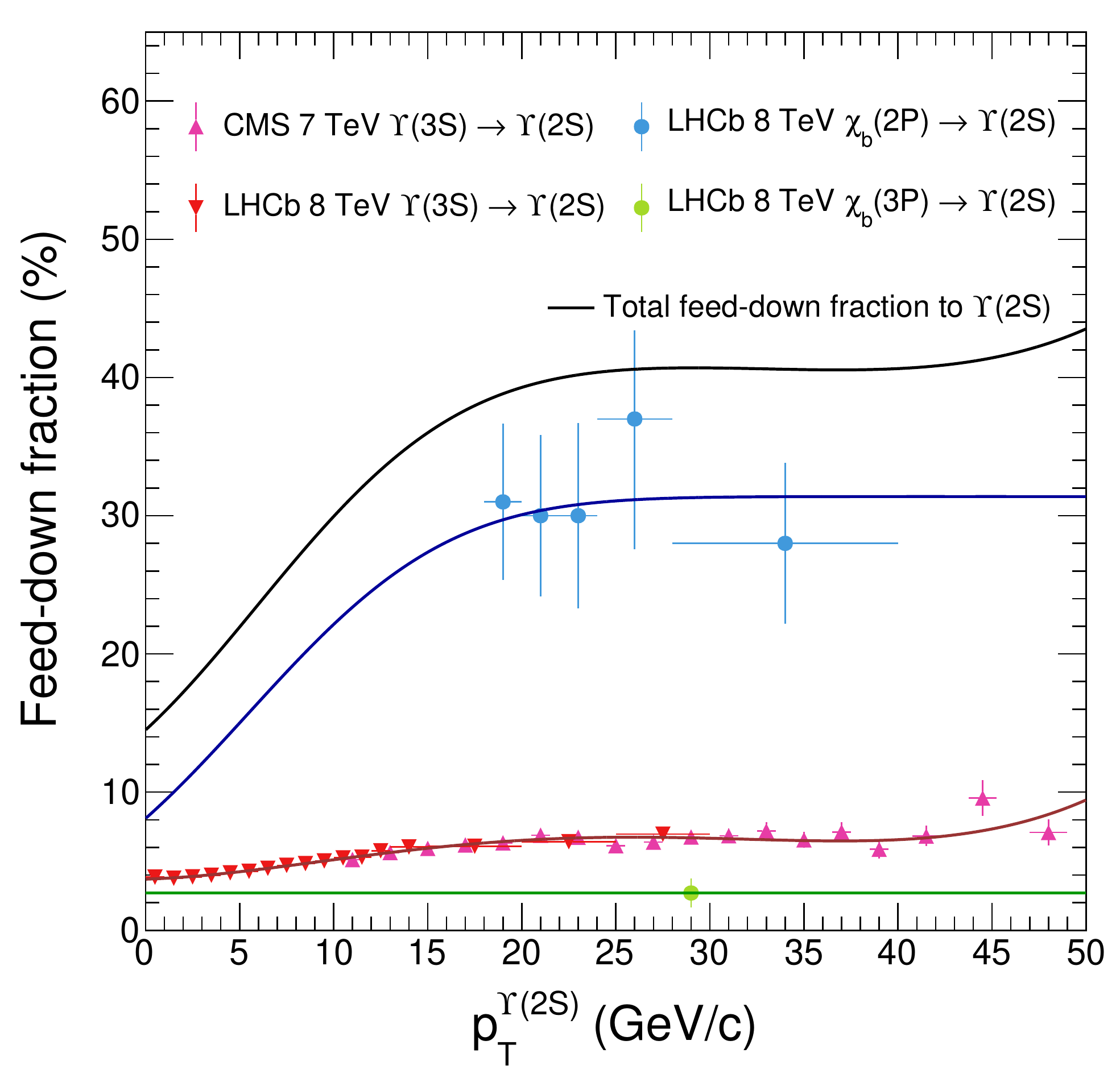}
        \includegraphics[width=0.8\linewidth]{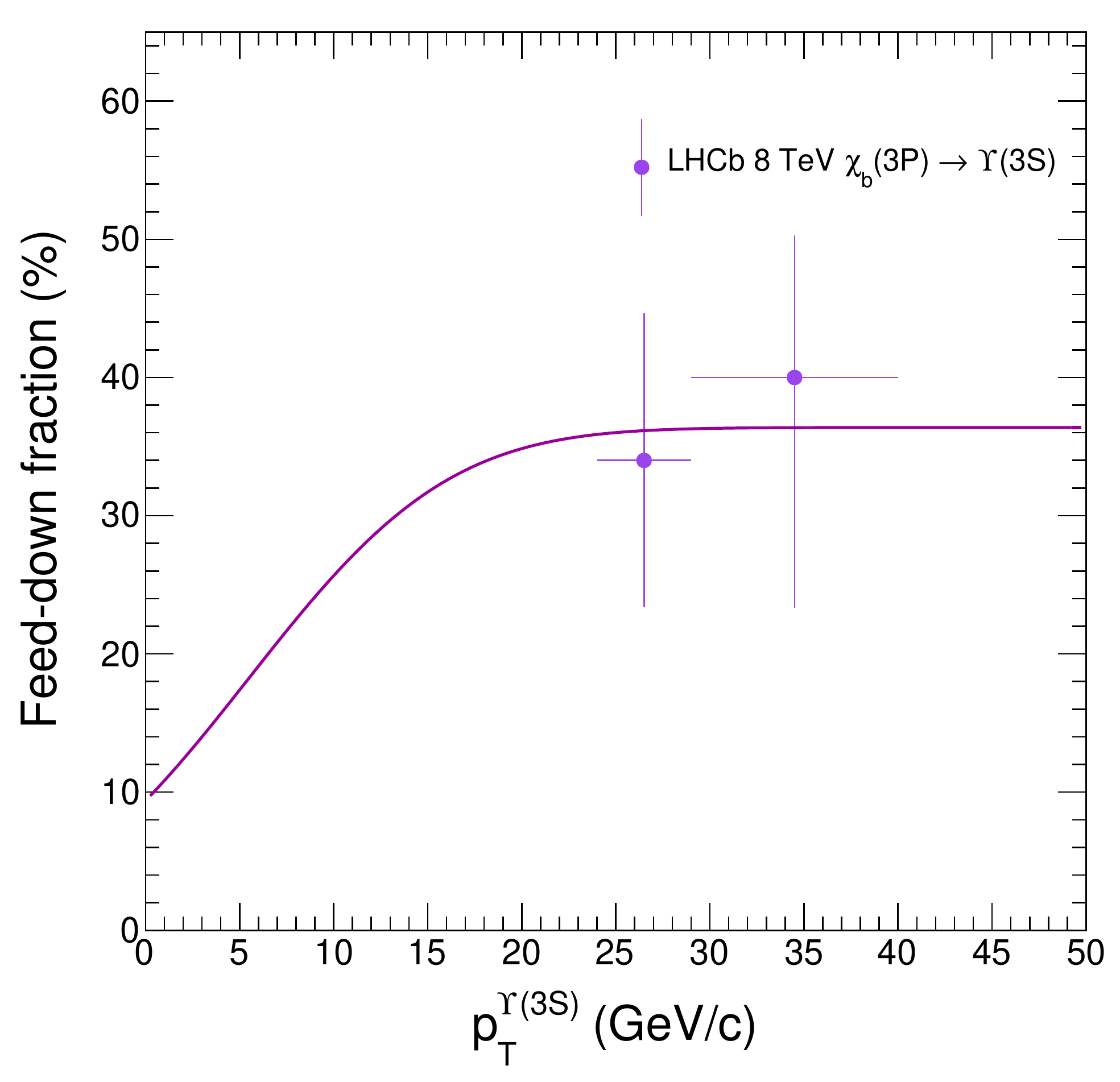}
    \caption{Feed-down fraction for \Ups 1 (top), \Ups 2 (middle), and \Ups 3 (bottom) states.}
    \label{fig:FeedDown}
\end{figure}

To simulate the medium response of inclusive bottomonium production, the  contributions from feed-down decays, i.e. decays from higher excited states, need to be carefully considered.
Here we consider only strong or electromagnetic decay modes of excited states in the feed-down decays, as the decays from $H$, $Z$, and $W$ bosons to $\Upsilon$ states have negligible effects to the inclusive yields. 
In general, the feed-down component from a $Q_{m}$ state to a $Q_{n}$ state ($Q_{m}$ being a higher excited state) can be calculated as 

\begin{equation}
\label{eq:FD}
    \mathcal{F}^{Q_{m}}_{Q_{n}} = \mathcal{B}(Q_{m}\rightarrow Q_{n})\frac{\sigma_{Q_{m}}}{\sigma_{Q_{n}}},
\end{equation}
where $\mathcal{B}$ denotes the branching ratio of $Q_{m}$ into $Q_{n}$ and $\sigma_{Q}$ referring the cross section of the corresponding state. 
We used publish experimental data in \pp collisions at LHC by CMS~\cite{CMS:2015FD} and LHCb~\cite{LHCb:2015FD,LHCb:2015FDchiB} to estimate the feed-down fraction. 
Figure~\ref{fig:FeedDown} shows the feed-down fraction from higher excited states for each \Ups 1, \Ups 2, and \Ups 3 states. 
The feed-down fractions from CMS are calculated using the presented cross section ratios multiplied by the PDG world-average branching ratios~\cite{Zyla:2020zbs} as in Eq.~\ref{eq:FD}.
The data points of each feed-down component are fitted with an empirical function and their sum drawn as the black solid line which is the total amount of feed-down fraction for the given $\Upsilon$ state.
For \Ups 2 and \Ups 3, the fit function is parameterized to be the same as for the P-wave state feed-down function of \Ups 1 because of the absence of experimental measurements at low-\pt. Afterwards, the function is normalized to match the data points at high-\pt for each \Ups 2 and \Ups 3 meson, respectively. 

The feed-down contribution for the nuclear modification factor and elliptic flow of inclusive \Ups n is considered by calculating an weighted average of the quantities for possible states as

\begin{align*}
    R_{n}(p_T) = \sum_{i} R_{i}(p_T)  \mathcal{F}^{Q_{i}}_{Q_{n}} (p_T),
\end{align*}
where $R_{n}$ is the weighted averaged value for a certain \Ups n state, $R_{i}$ is the value for a certain state contributing the the \Ups n state, and $\mathcal{F}^{Q_{i}}_{Q_{n}}$ is the feed-down fraction. The effect of possible \pt shifts by the decays are ignored as the daughter particle carries mostly the mother $\Upsilon$'s momentum because of their similar masses. Since we do not have the nuclear modification factor and elliptic flow for $\chi_b$ states, it is assumed that $R_{\Upsilon(2S)} \approx R_{\chi_{b}(1P)}$ and $R_{\Upsilon(3S)} \approx R_{\chi_{b}(2P)} \approx R_{\chi_{b}(3P)}$ like the study in Ref.~\cite{Hong:2019ade}.

\clearpage

\section{Results and Discussions}
\label{sec:Results_Disscussion}

This section presents and discuss the obtained results of nuclear modification factors (\rpa, \raa) and elliptic flow (\vtwo) for \Ups 1, \Ups 2, and \Ups 3 mesons in \pPb, \pO, and \OO collisions at \sqsn = 8 TeV.
The performance of our framework is tested in nucleus-nucleus collisions and compared with the experimental data at LHC as described in Sec.~\ref{sec:res_aa}. 
The results of the nuclear modification factors and \vtwo in the three small collision systems are shown and discussed in Sec.~\ref{sec:res_rpa} and Sec.~\ref{sec:res_vtwo}, respectively.
For all results, the uncertainty represents the statistical uncertainty that arises from the number of generated events. Also, the feed-down corrections are applied in all calculations as described in Sec.~\ref{sec:FDCorr}.

\subsection{Framework demonstration in \PbPb }
\label{sec:res_aa}
\begin{figure}[htb]
    \centering
        \includegraphics[width=0.9\linewidth]{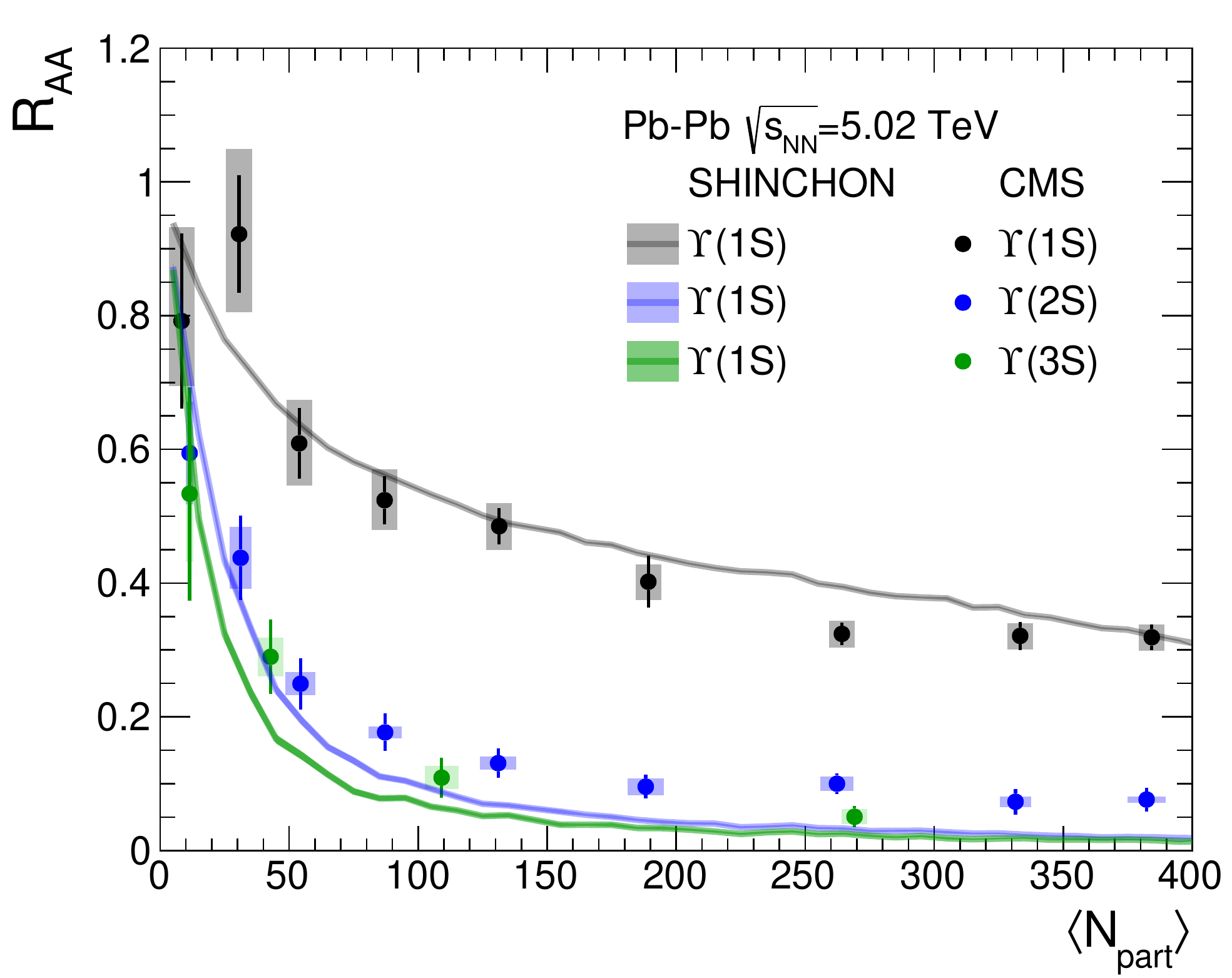}
    \caption{Calculated \raa for $\Upsilon$ states as a function of $\langle \Npart \rangle$ in \PbPb collisions at \sqsn = 5.02 TeV. The uncertainties of the curves represent the statistical uncertainty to the corresponding simulated events. The data points are taken from results by CMS~\cite{CMS:2018zza,CMS-PAS-HIN-21-007}.}
    \label{fig:raa_npart_pbpb}
\end{figure}

The medium response of the full simulation is demonstrated in \PbPb collisions at \sqsn = 5.02 TeV. Figure~\ref{fig:raa_npart_pbpb} shows the \raa curves for \Ups 1, \Ups 2, and \Ups 3 as a function of $\langle \Npart \rangle$ in \PbPb collision at \snn = 5.02 TeV together with the measurements from CMS~\cite{CMS:2018zza,CMS-PAS-HIN-21-007}.  
The calculations agree well with the experimental data for \Ups 1 after applying feed-down corrections, while this correction seems to have restricted effects for \Ups 2 and \Ups 3. The results of the excited $\Upsilon$ states show consistency with data in peripheral collisions, although deviations are found towards central \PbPb collisions. 
The discrepancy in central collisions might be present due to the exclusion of recombination processes in this paper, which is expected to be more prominent at large $\langle \Npart \rangle$ values. 

\begin{figure}[htb]
    \centering
        \includegraphics[width=0.9\linewidth]{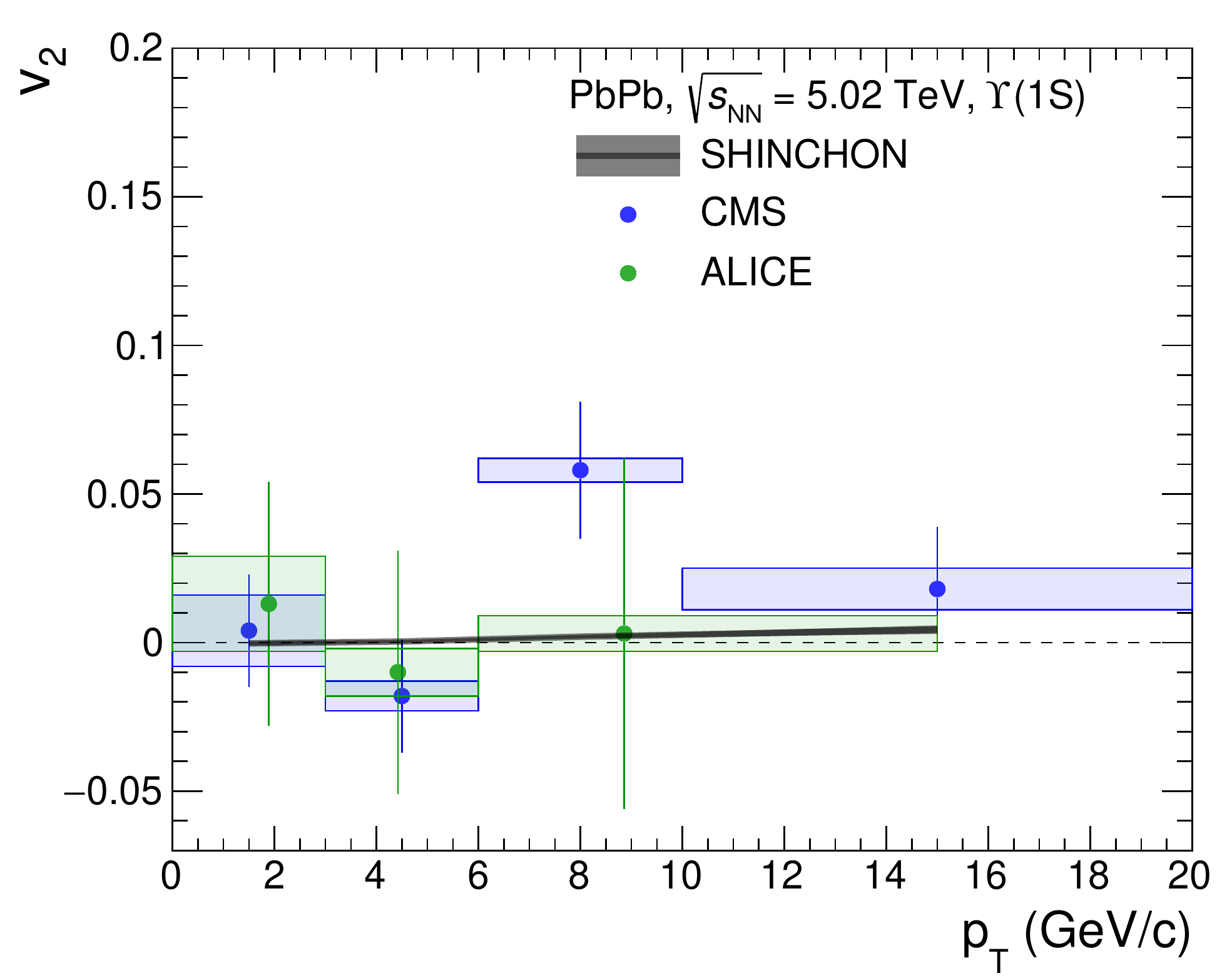}
    \caption{Calculated \vtwo for \Ups 1 as a function of \pt in \PbPb collsions at \sqsn = 5.02 TeV. The uncertainties of the curves represent the statistical uncertainty to the corresponding simulated events. The data points are taken from results by ALICE~\cite{ALICE:2019Yv2} and CMS~\cite{CMS:2021Yv2}.}
    \label{fig:v2_aa_pt}
\end{figure}

Figure~\ref{fig:v2_aa_pt} shows the computed results of \vtwo as a function of \pt for \Ups 1 mesons under the same conditions for the \PbPb \raa calculations. The results indicate very small \vtwo values in \PbPb collisions for \Ups 1 mesons which is consistent with the measurements from ALICE~\cite{ALICE:2019Yv2} and CMS~\cite{CMS:2021Yv2}. 

\subsection{Nuclear modification factor}
\label{sec:res_rpa}

\begin{figure}[htb]
    \centering
        \includegraphics[width=0.8\linewidth]{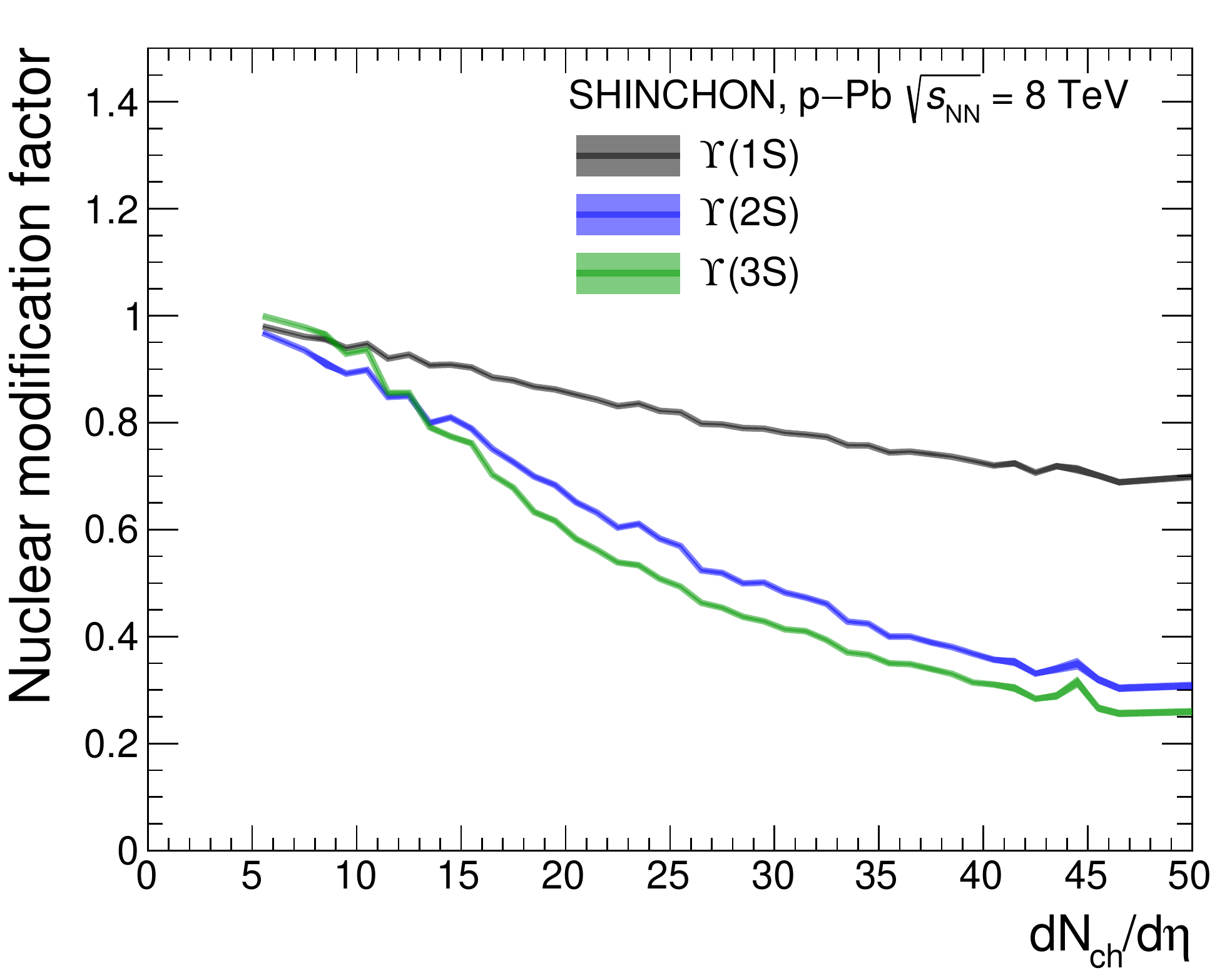}
        \includegraphics[width=0.8\linewidth]{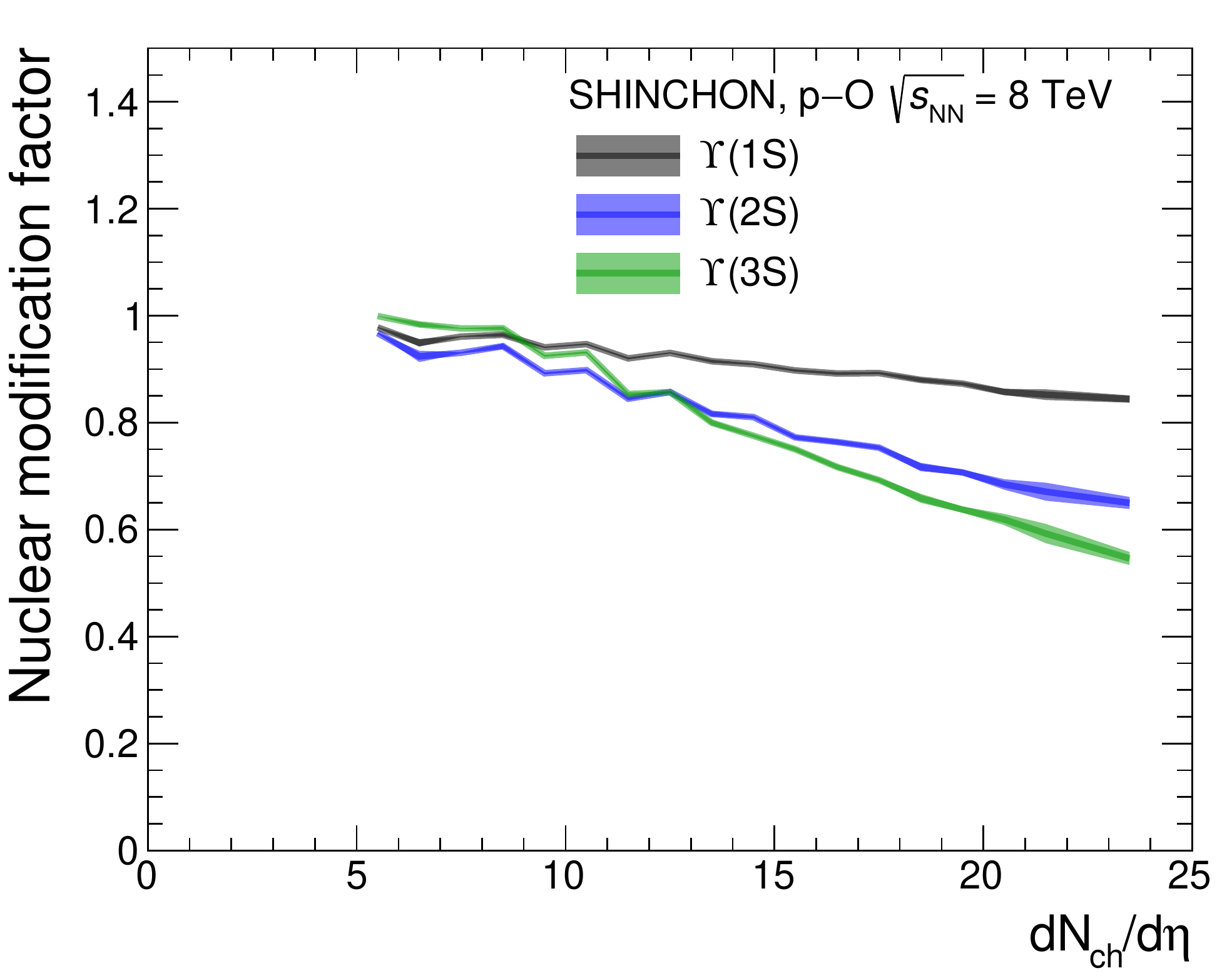}
        \includegraphics[width=0.8\linewidth]{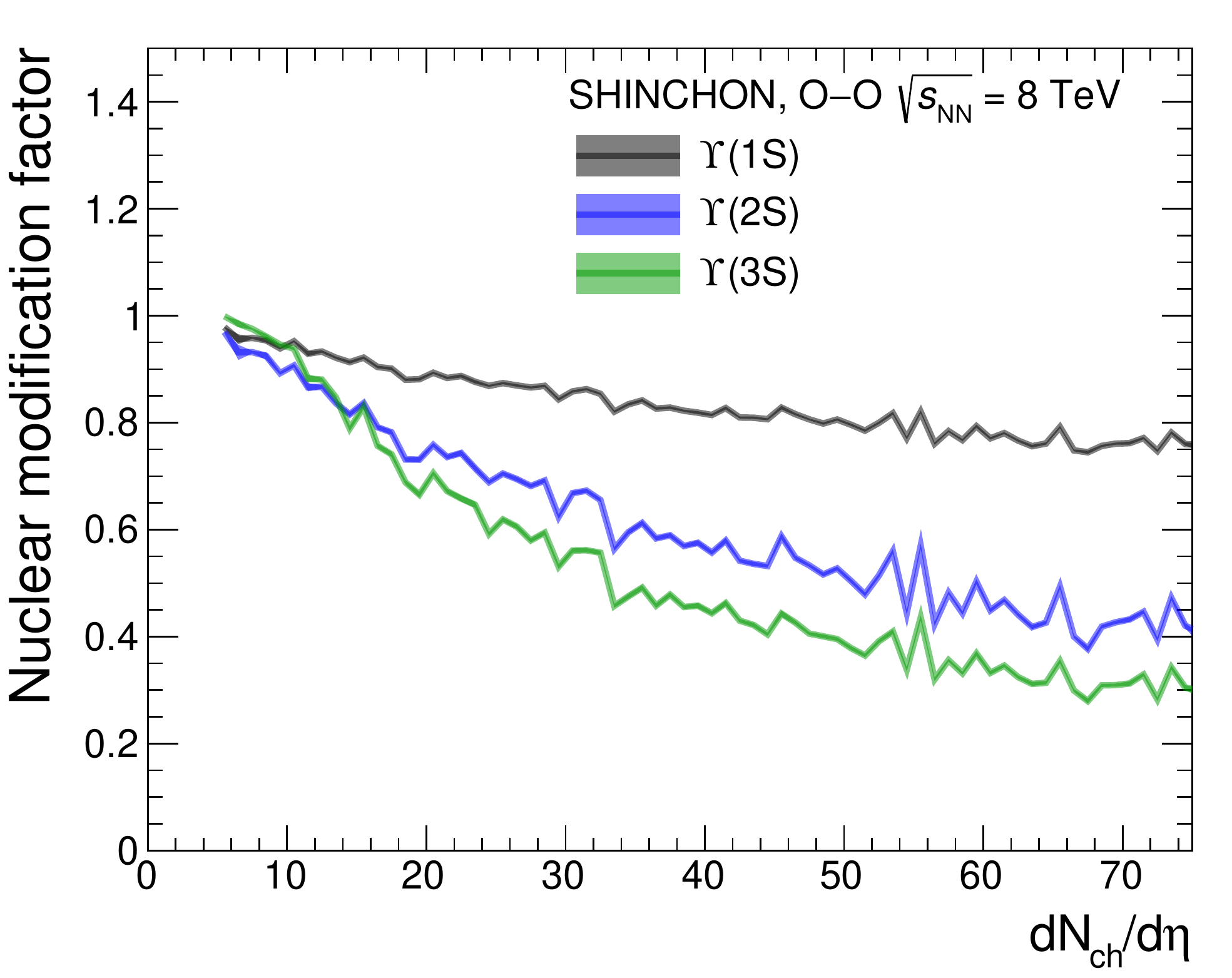}
    \caption{Nuclear modification factors for \Ups 1, \Ups 2, and \Ups 3 as a function of $dN_{ch}/d\eta$ in \pPb (top), \pO (middle), and \OO (bottom) collisions at \sqsn = 8 TeV. The uncertainties represent the statistical uncertainty to the corresponding simulated events.}
    \label{fig:rpa_mult}
\end{figure}

The calculated nuclear modification factor as a function of $dN_{ch}/d\eta$ for \Ups 1, \Ups 2, and \Ups 3 mesons in \pPb, \pO, and \OO collisions at \sqsn = 8 TeV are shown in Fig.~\ref{fig:rpa_mult}. 
The nuclear modification factor shows a gradual decrease with increasing event multiplicity for all three $\Upsilon$ states in all three collision systems. Also, the amount of suppression towards higher multiplicity events is found to be sequentially ordered following the magnitude of the corresponding $\Upsilon$ binding energy.

\begin{figure}[htb]
    \centering
        \includegraphics[width=0.8\linewidth]{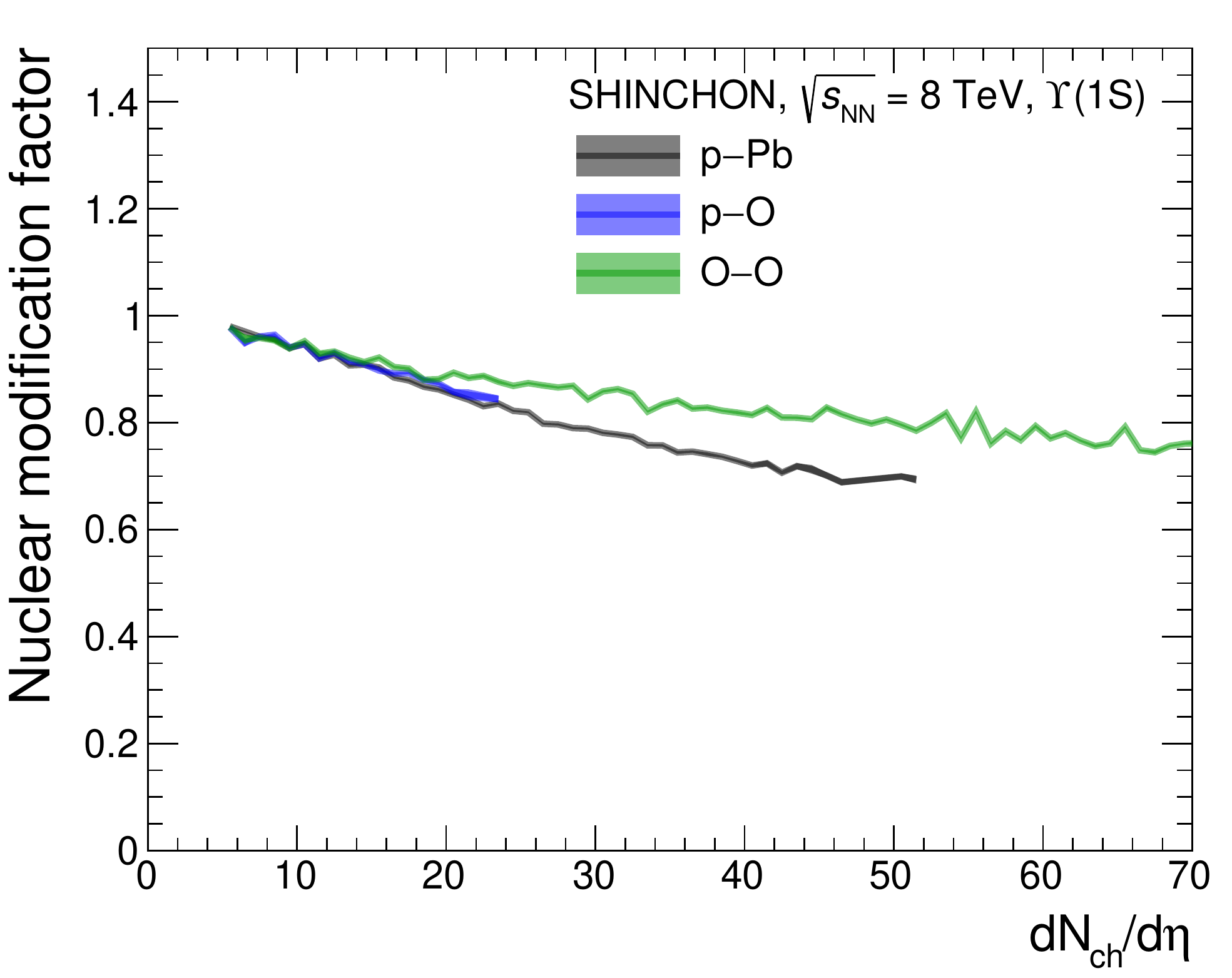}
        \includegraphics[width=0.8\linewidth]{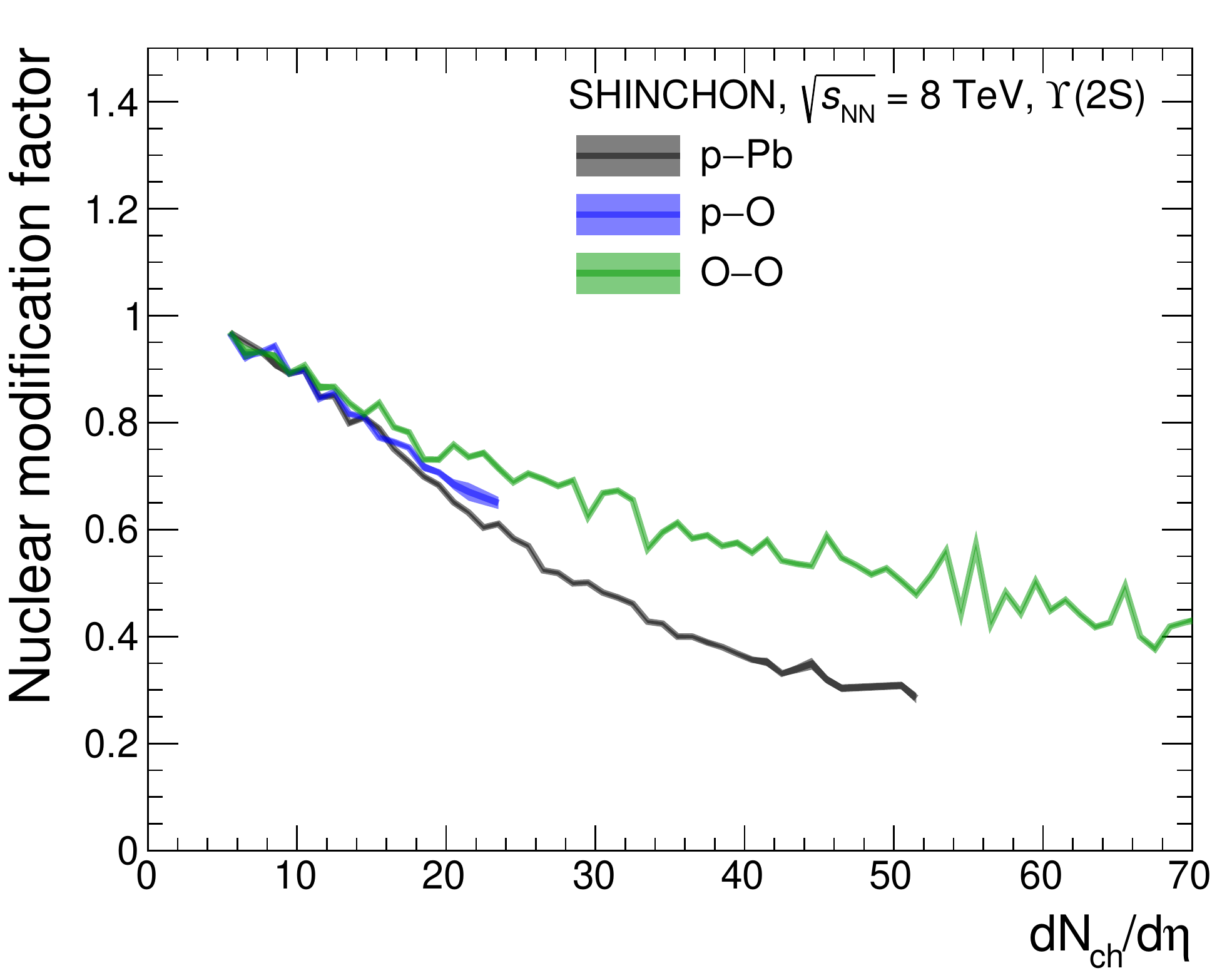}
        \includegraphics[width=0.8\linewidth]{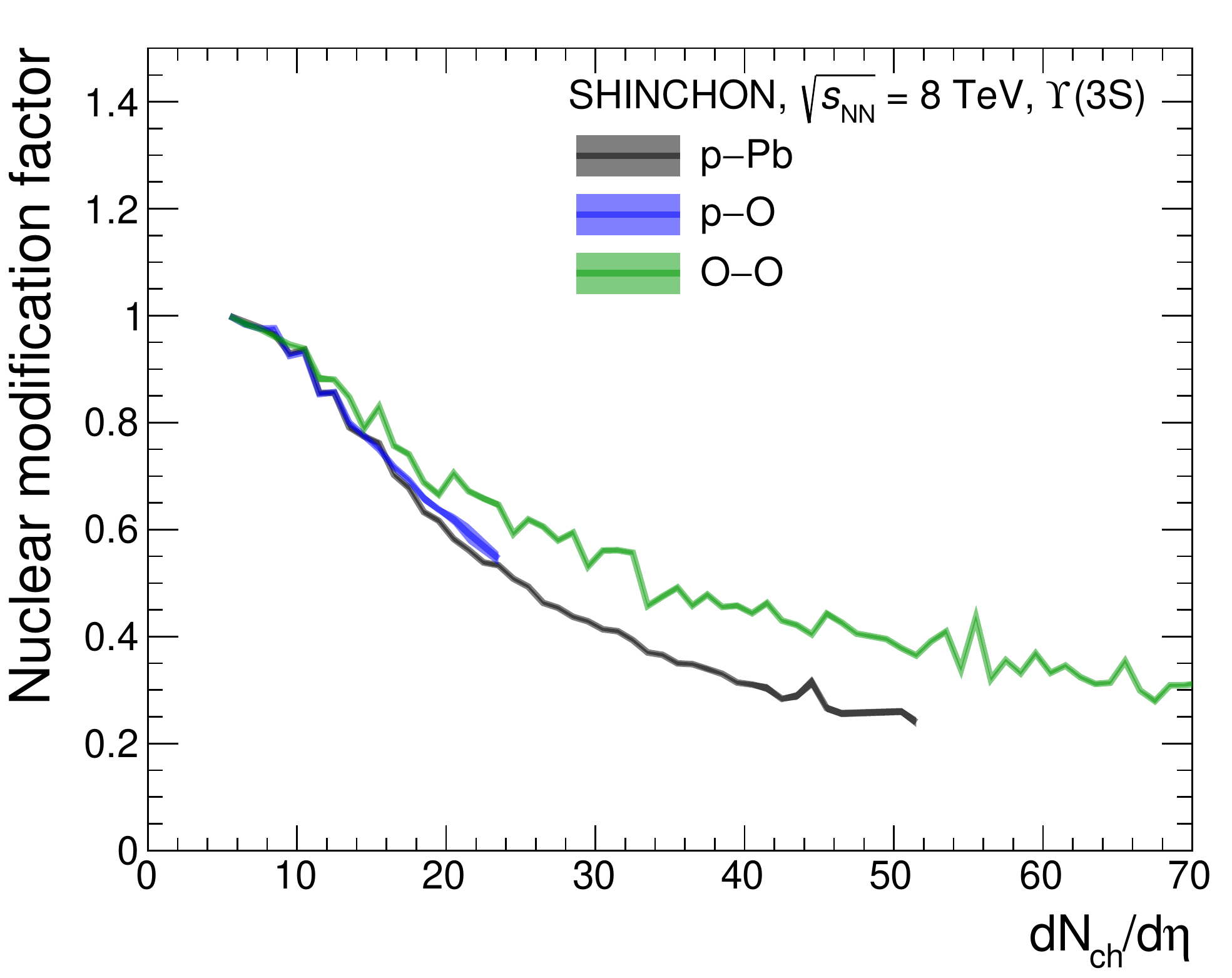}
    \caption{Calculated nuclear modification factor as a function of $dN_{ch}/d\eta$ in \pPb, \pO, and \OO collisions at \sqsn = 8 TeV each for \Ups 1 (top), \Ups 2 (middle), \Ups 3 (bottom). The uncertainties represent the statistical uncertainty to the corresponding simulated events.}
    \label{fig:rpa_mult_coll}
\end{figure}

To diagnose the strength of modification among different collision systems, the nuclear modification factors are separately displayed for each $\Upsilon$ state as shown in Fig.~\ref{fig:rpa_mult_coll}.
In the low multiplicity region ($dN_{ch}/d\eta < 25)$, the suppression level of $\Upsilon$ states is found to be similar in \pPb and \pO collisions, while a slightly less suppression is seen in \OO collisions in particular for \Ups 2 and \Ups 3. 
Despite the medium size in \OO collisions being the largest and that in \pO collisions the smallest, the amount of suppression is not proportional to the size of the created system. 
The fact that the system size is larger in a given multiplicity event rather implies a smaller energy density of the medium. Therefore, we conclude that the weaker suppression in \OO collisions is accounted for the smaller energy density compared to \pO and \pPb collisions.
At higher multiplicity region, where the results in \pPb and \OO collisions are compared, $\Upsilon$ states are stronger suppressed in \pPb collisions than in \OO collisions which is consistent with the finding at low multiplicity because of the same reason. 

\begin{figure}[htb]
    \centering
        \includegraphics[width=0.8\linewidth]{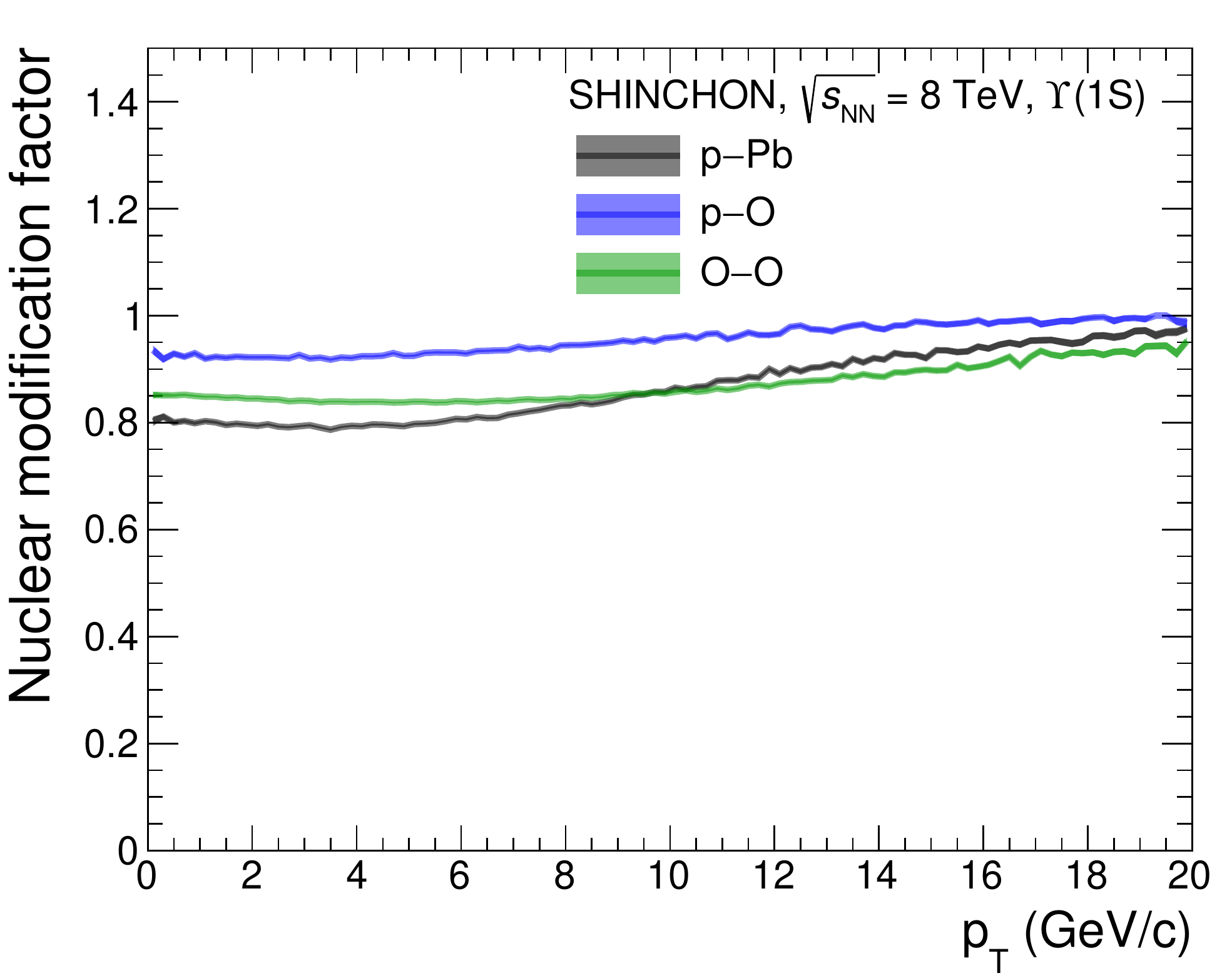}
        \includegraphics[width=0.8\linewidth]{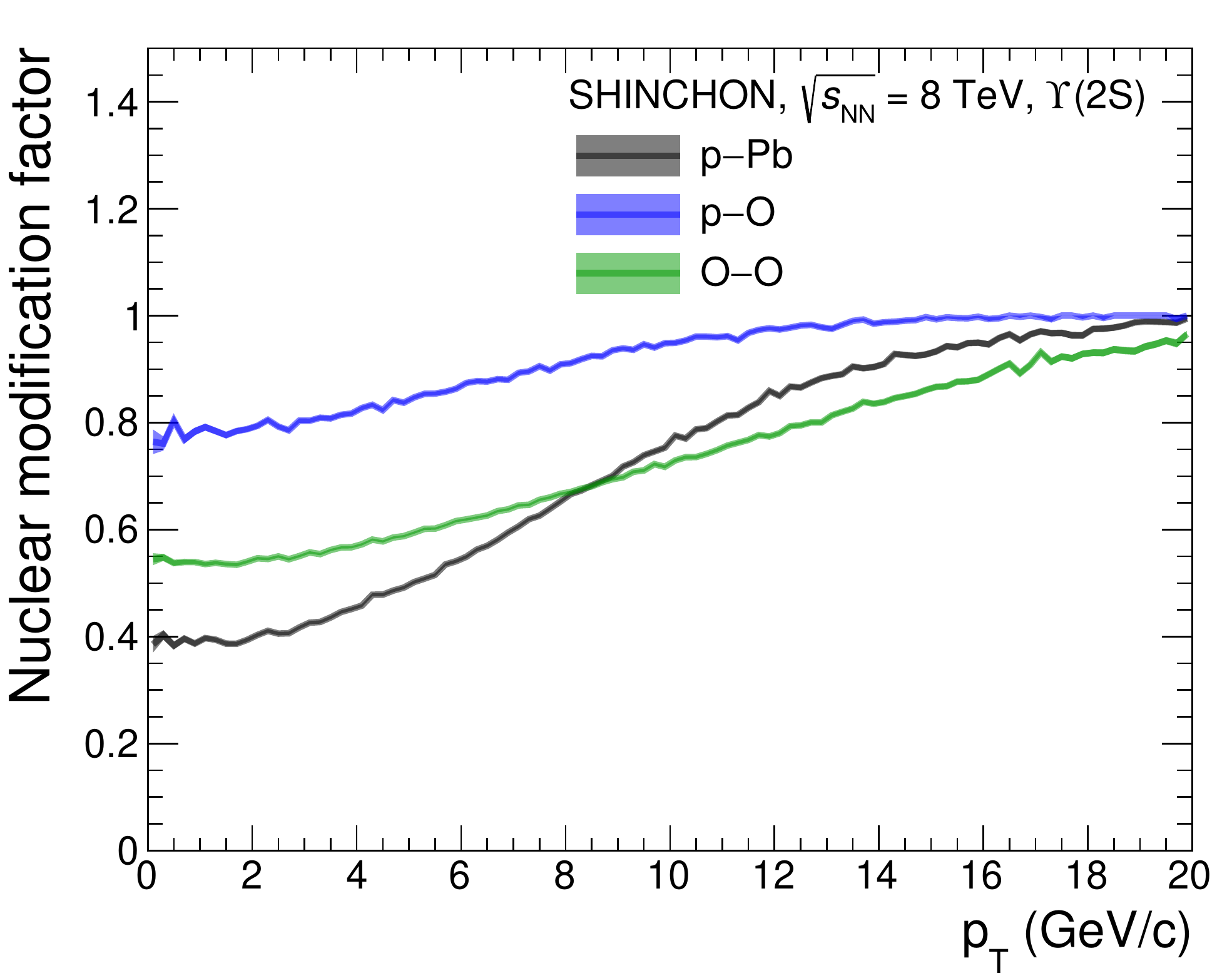}
        \includegraphics[width=0.8\linewidth]{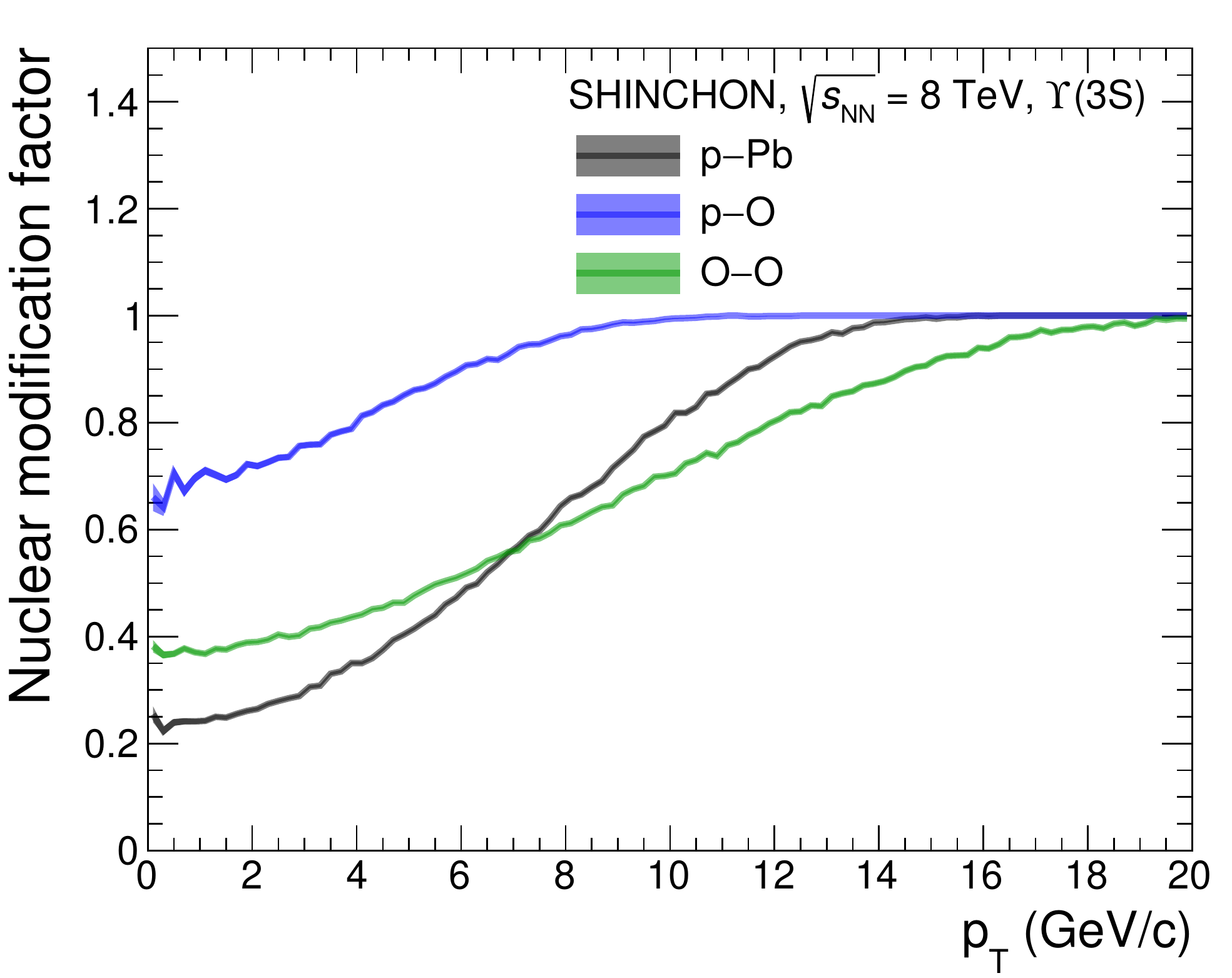}
    \caption{Nuclear modification factors for \Ups 1, \Ups 2, and \Ups 3 as a function of \pt in \pO, \pPb, and \OO collisions at \sqsn = 8 TeV. The uncertainties represent the statistical uncertainty to the corresponding simulated events.}
    \label{fig:rpa_pt}
\end{figure}

Figure~\ref{fig:rpa_pt} shows the same quantities presented as a function of \pt. The ordering of the nuclear modification factors is similar at low-\pt as in Fig.~\ref{fig:rpa_mult_coll}, but found to be reversed between \pPb and \OO collisions at high-\pt. 
Since the formation time of the $\Upsilon$ states is delayed towards the higher \pt region, the effective interaction time with the medium is also reduced. Such effect is more prominent in \pPb collisions because of the smaller initial medium size. 
The relation between the medium size and energy density affects the $\Upsilon$ suppression in the opposite direction for low- and high-\pt.
These results show that the shape of the \pt-dependent nuclear modification factor is sensitive to the formation time and initial collision geometry. 

\begin{figure}[htb]
    \centering
        \includegraphics[width=0.8\linewidth]{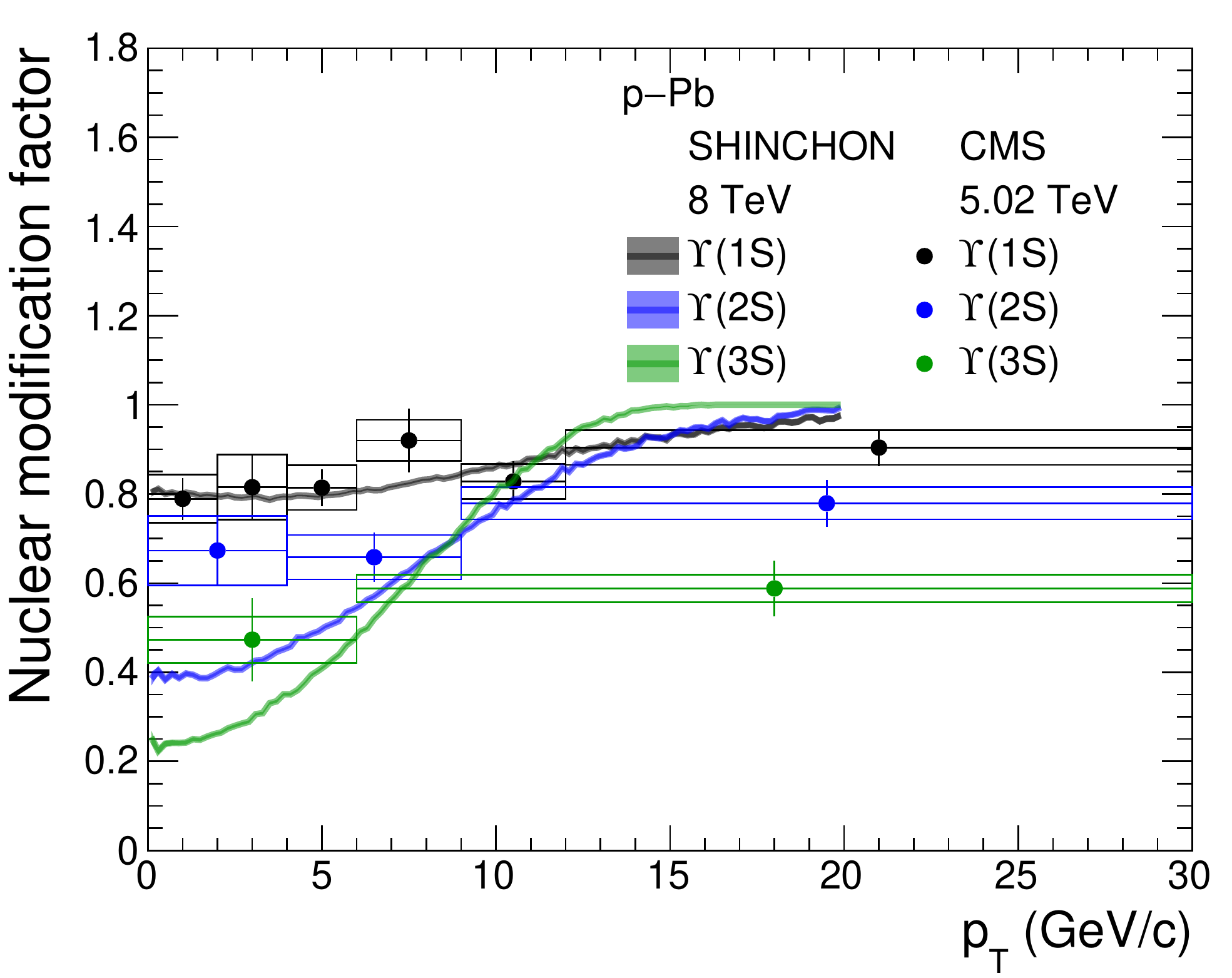}
    \caption{Calculated nuclear modification factor for \Ups 1, \Ups 2, and \Ups 3 as a function of \pt in \pPb  collisions at \sqsn = 8 TeV compared with the CMS results at \sqsn = 5.02 TeV.}
    \label{fig:rpa_pt_cms}
\end{figure}

We also compare our calculations with measured data by CMS in \pPb collisions at \sqsn = 5.02 TeV~\cite{CMS:2022wfi} as shown in Fig.~\ref{fig:rpa_pt_cms}. 
The results are in good agreement with data for \Ups 1, whereas deviations are seen for \Ups 2 and \Ups 3. 
Note that our \rpa values are for 8 TeV, in which the average multiplicity is about 15\% higher than that at 5.02 TeV~\cite{CMS:2017shj}. This difference affects the modification of $\Upsilon$ yields more significantly at low-\pt, where they mostly experience the full medium evolution.
Although the mean \pt of the highest \pt bin of the CMS measurements for the \Ups 2 and \Ups 3 meson is expected to be slightly lower than the center of the bin, our calculations still overshoot the experimental data. 
As discussed in Fig.~\ref{fig:rpa_pt}, this disagreement is possibly related to the formation time and initial collision geometry.
Note that initial-state model calculations considering modification of nuclear parton distribution functions and energy loss show about 10\% suppression of \Ups 1 at mid-rapidity~\cite{CMS:2017shj,Vogt:2015uba,Arleo:2014oha}.

\subsection{Elliptic flow}
\label{sec:res_vtwo}

\begin{figure}[htb]
    \centering
        \includegraphics[width=0.8\linewidth]{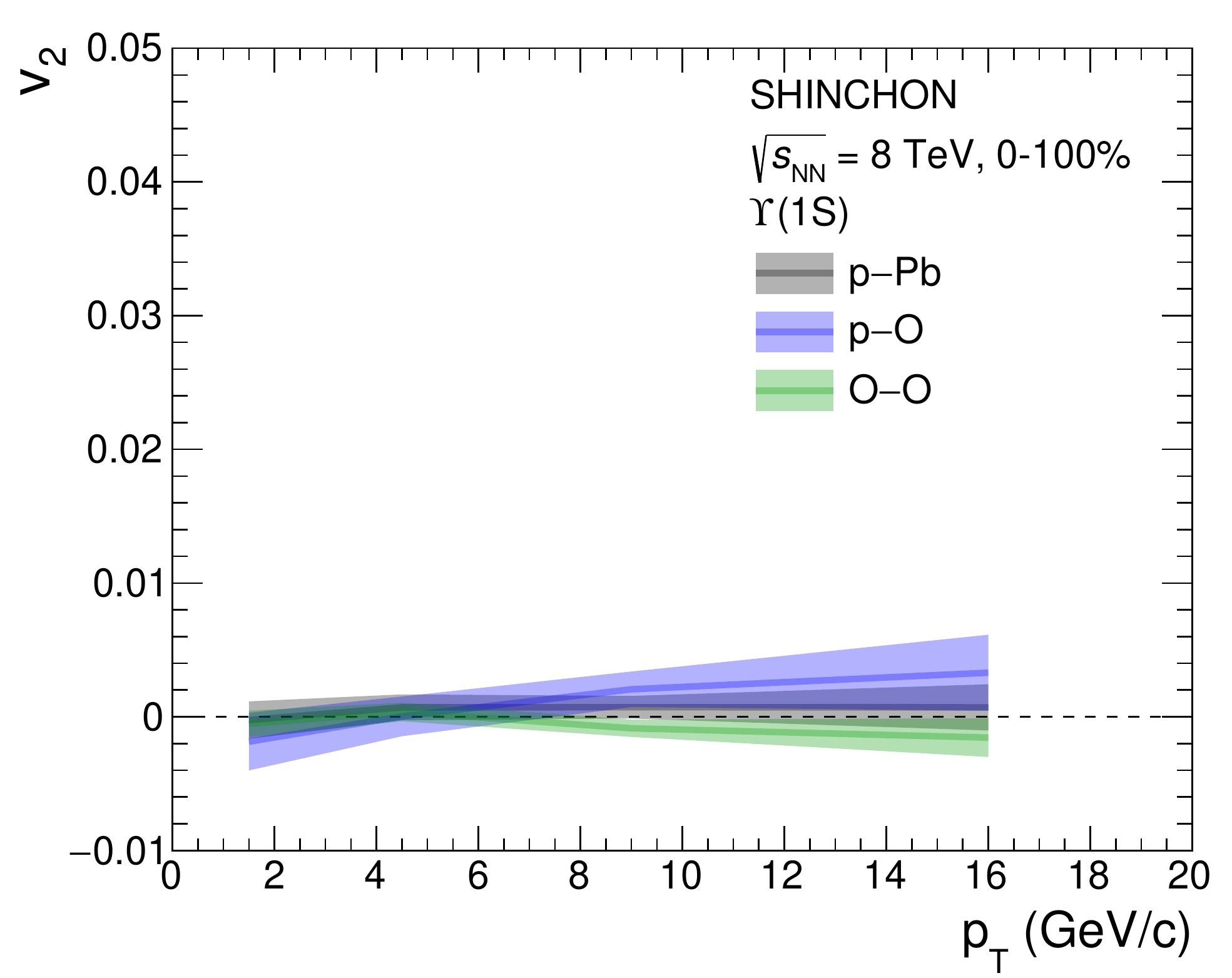}
        \includegraphics[width=0.8\linewidth]{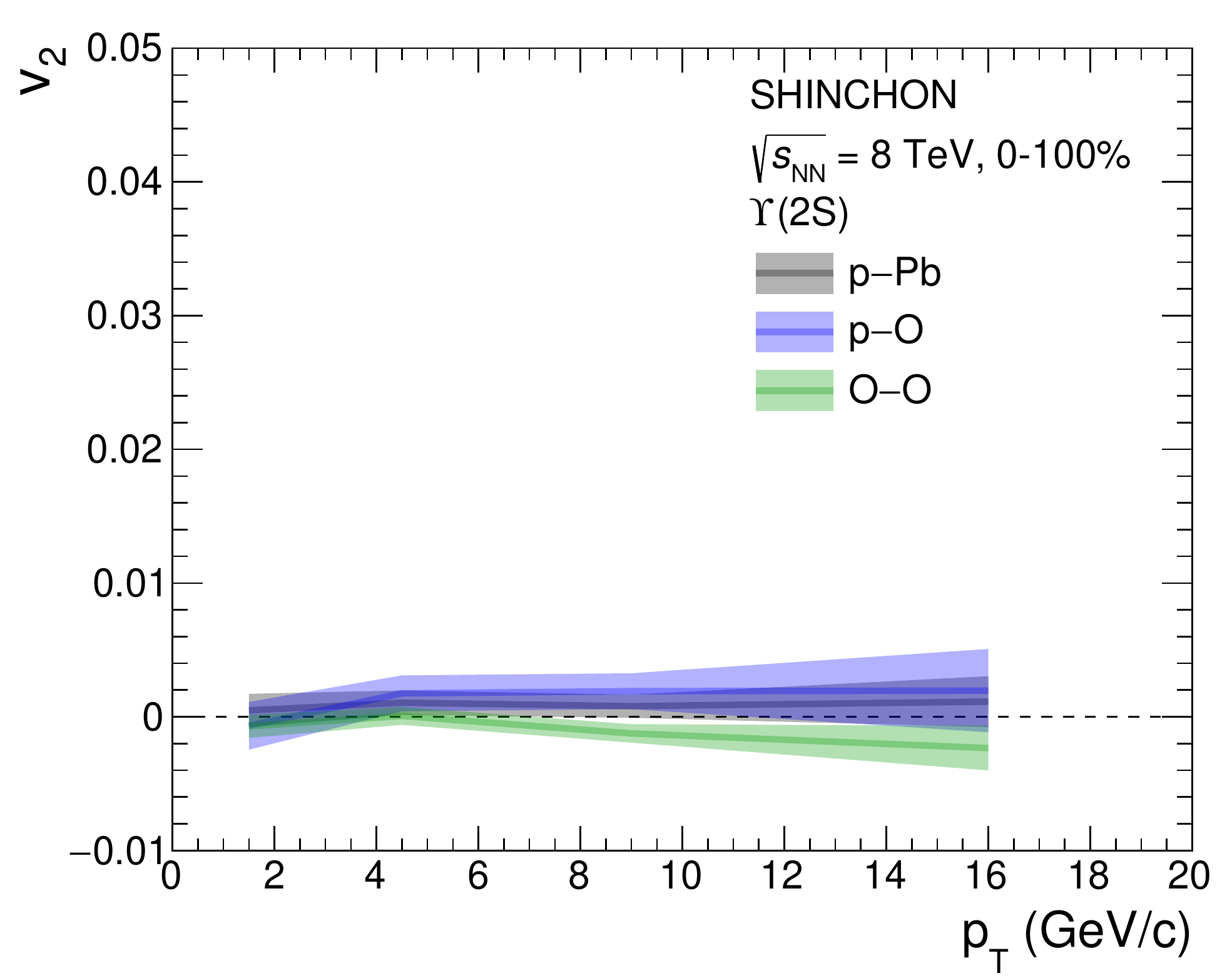}
        \includegraphics[width=0.8\linewidth]{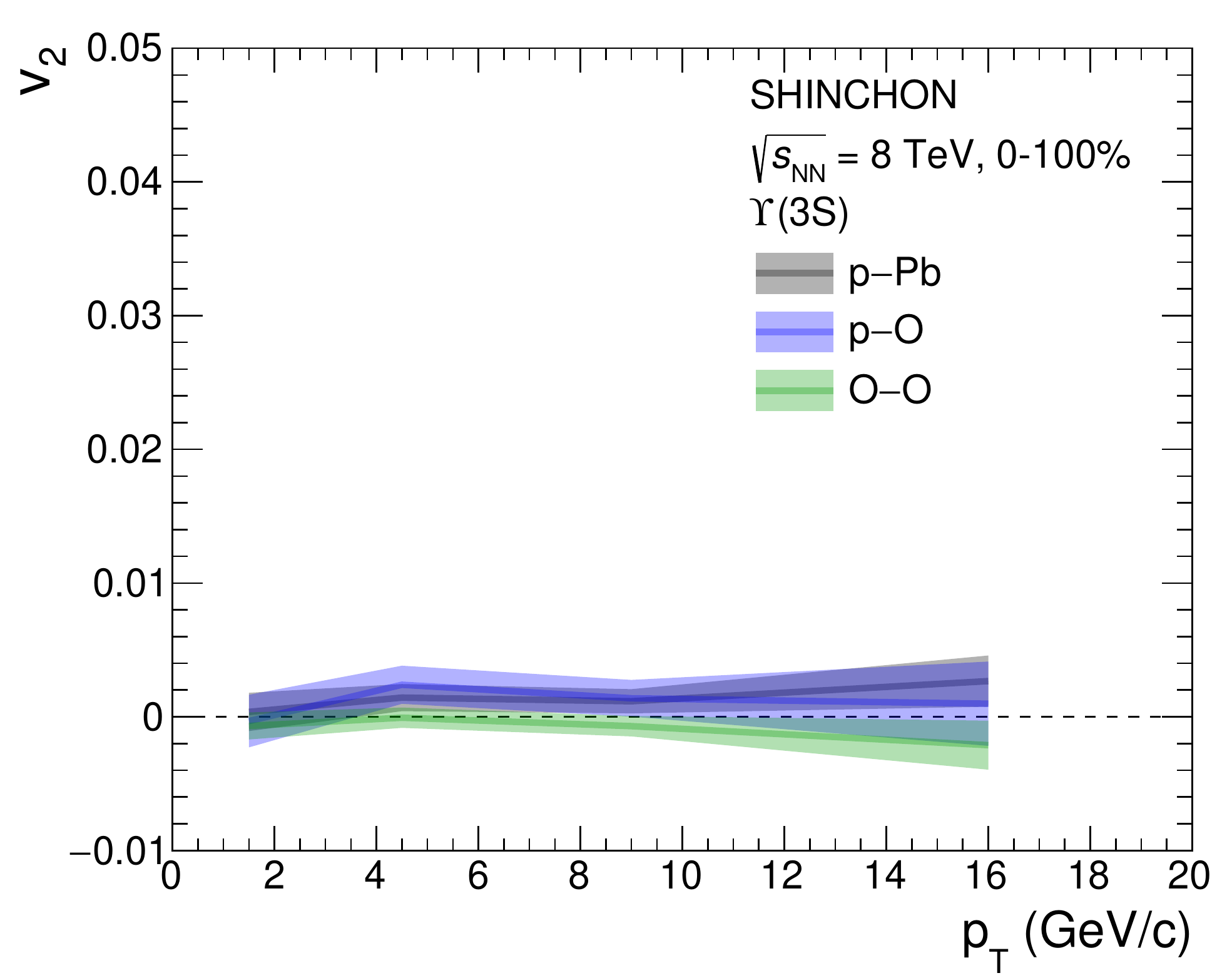}
    \caption{Calculated $v_2$ in \pPb, \pO, and \OO collisions as a function of \pt for \Ups 1 (top), \Ups 2 (middle), \Ups 3 (bottom) at \sqsn = 8 TeV. The uncertainties represent the statistical uncertainty to the corresponding simulated events.}
    \label{fig:v2_pt}
\end{figure}

The \vtwo is calculated in the same setting as in the calculation for the nuclear modification factors in Sec.~\ref{sec:res_rpa} and shown in Fig.~\ref{fig:v2_pt}. For all three $\Upsilon$ states, the \vtwo values are consistent with zero in the overall \pt region in \pPb, \pO, and \OO collisions.
Also, their \vtwo values are very similar among the three collision systems. It may indicate that the elongated formation time of $\Upsilon$s towards high-\pt is not as significant as implemented in the employed model.
To test the relation between \vtwo and the amount of suppression, we computed the \vtwo in the 0--5\% selected high-multiplicity events. 
For all $\Upsilon$ states, the \vtwo values are found to be larger at high-multiplicity only with a small magnitude. 
\begin{figure}[htb]
    \centering
        \includegraphics[width=0.8\linewidth]{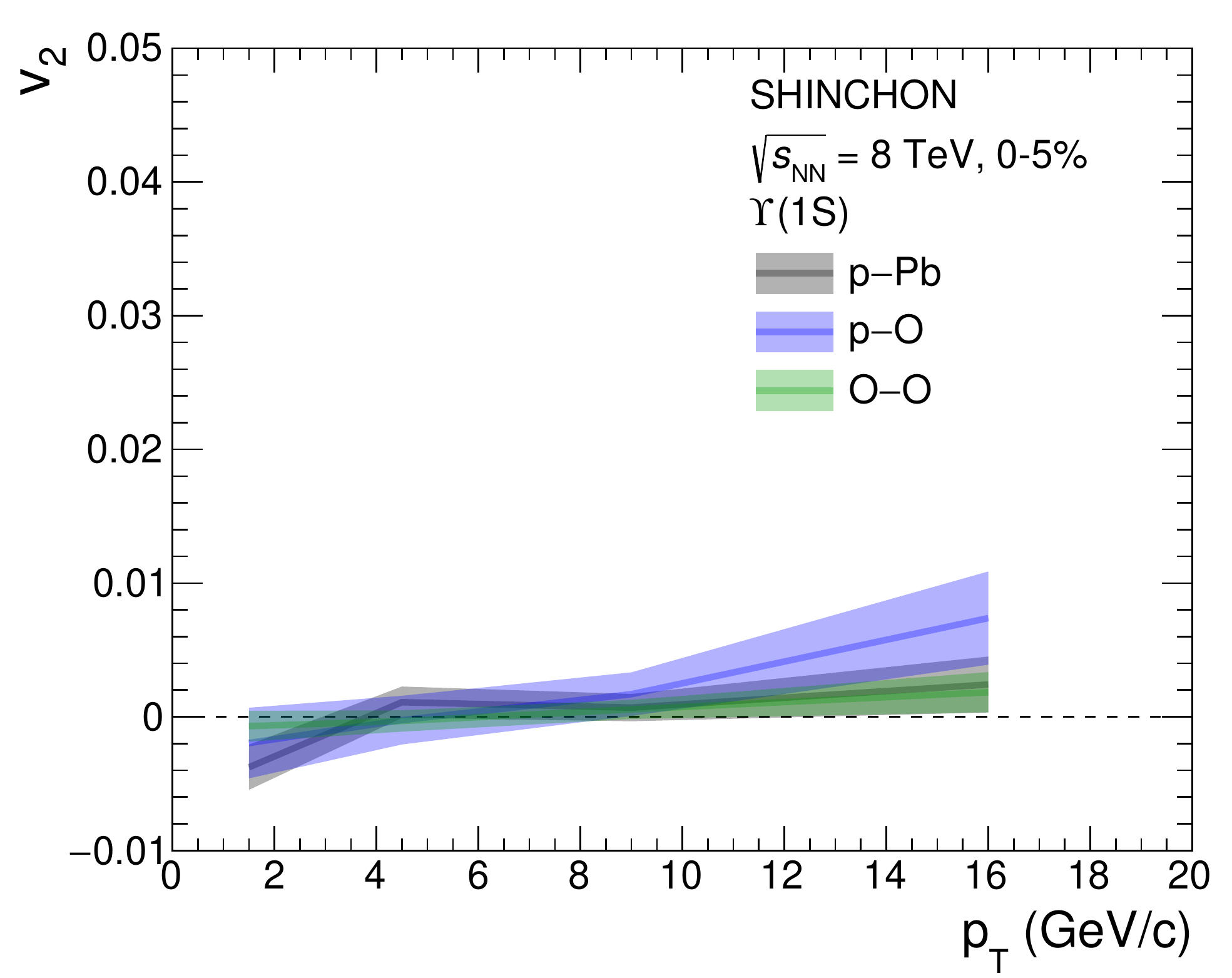}
        \includegraphics[width=0.8\linewidth]{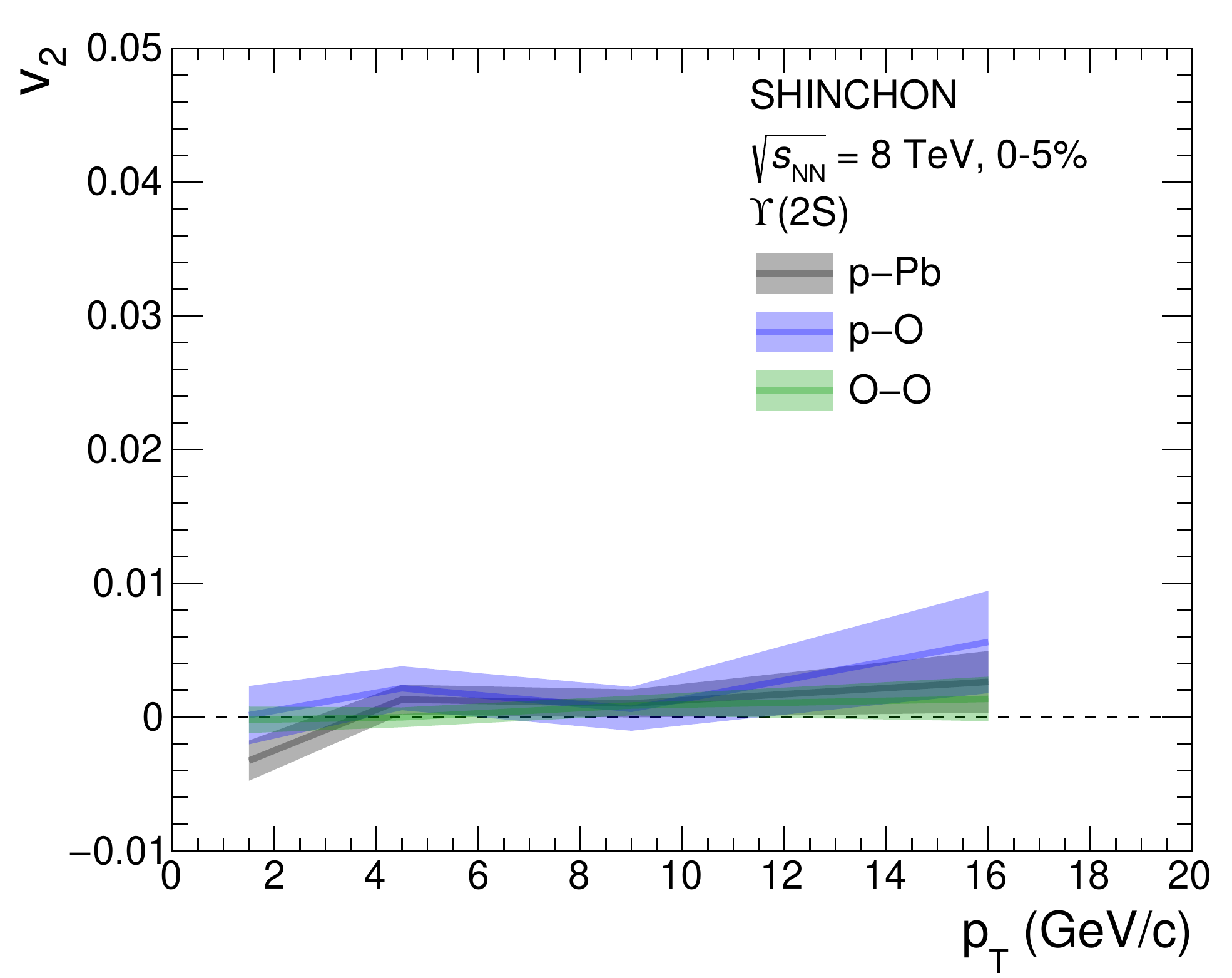}
        \includegraphics[width=0.8\linewidth]{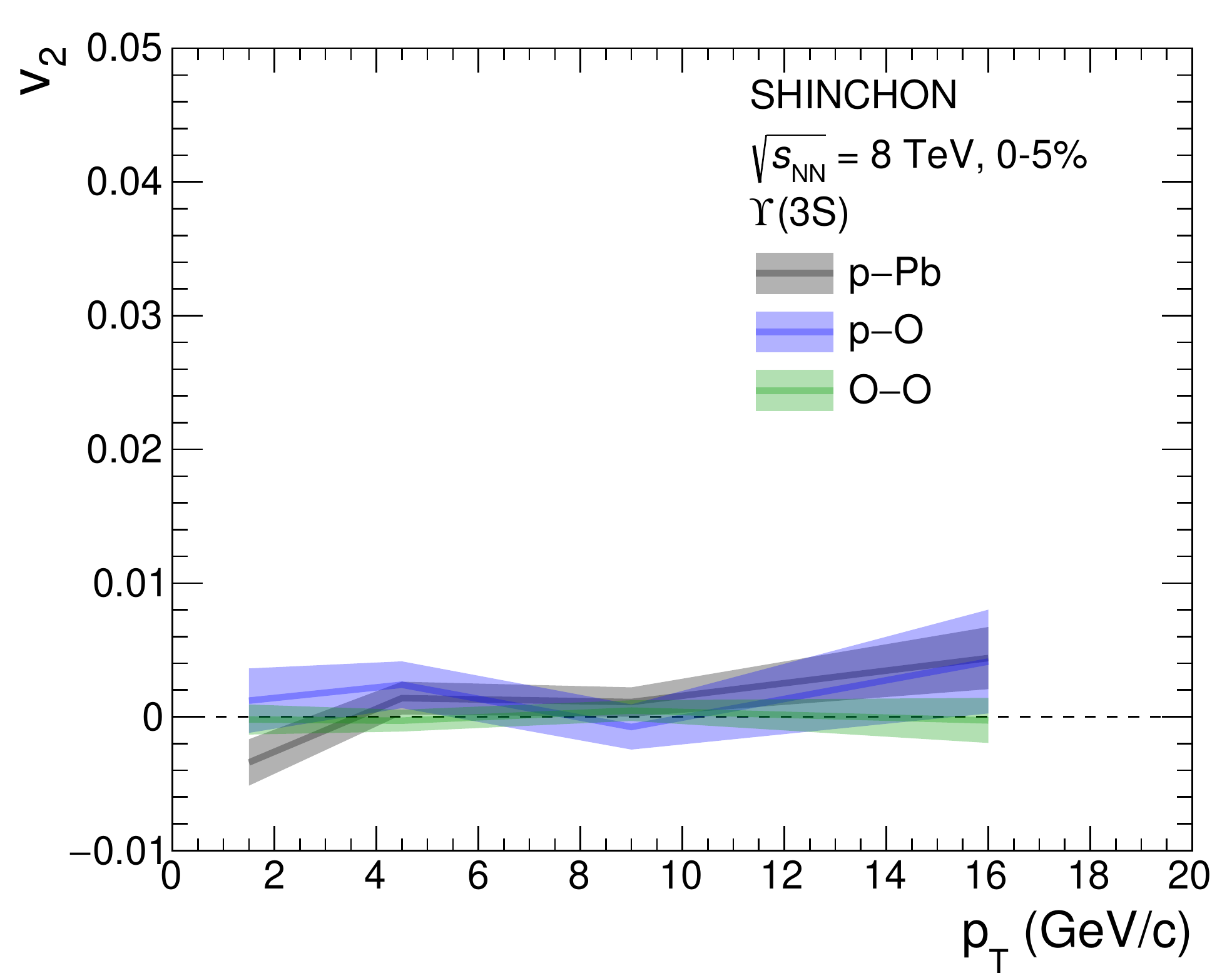}
    \caption{Calculated $v_2$ in \pPb, \pO, and \OO collisions as a function of \pt for \Ups 1 (top), \Ups 2 (middle), \Ups 3 (bottom) at \sqsn = 8 TeV in 0--5\% high multiplicity events. The uncertainties represent the statistical uncertainty to the corresponding simulated events}
    \label{fig:v2_pt_hm}
\end{figure}

Although the yields of $\Upsilon$ mesons are more strongly suppressed for excited states as well as for higher multiplicity events, their \vtwo values show an overall consistency with each other in all studied cases. 
In particular, since the low-\pt $\Upsilon$ mesons are traversing very slowly, it is unlikely expected that they escape the medium before it reaches the chemical freeze-out temperature. Therefore, such $\Upsilon$ mesons are not able to capture the anisotropy of the initial collision geometry despite their significant suppression. It is understood that the path-length dependent suppression could lead to a non-zero \vtwo which can be probed at high-\pt for $\Upsilon$ mesons. However, with our current simulation settings, we did not observe any firm non-zero \vtwo that deviates from that at low-\pt in small collision systems for all $\Upsilon$ states.

\pagebreak

\section{Summary}
\label{sec:Sum}

We have performed a Monte Carlo simulation study for the medium response of bottomonia in \pA and $A$+$A$ collisions. The simulation framework has been developed based on the theoretical calculation of the thermal width of \Ups n~\cite{Hong:2019ade} and the publicly available codes to describe the initial condition and evolution of heavy-ion collisions. 
In this initial work, only dissociation effect is considered, and we also considered the contribution of feed-down from higher excited states. To demonstrate the framework, we calculate the nuclear modification factor and elliptic flow of \Ups n and compare to experimental results. 
In the nuclear modification factor as a function of the number of participants, the model and data are comparable for \Ups 1 at the entire region, whereas the suppression for \Ups 2 and \Ups 3 are stronger in the model. 
In addition to the formation time, which could directly affect the magnitude of nuclear modification, the comparison for excited states would be improved when incorporating the regeneration effect. 
In case of the elliptic flow, the model expect $v_2<0.01$ which agrees with the experimental results~\cite{ALICE:2019Yv2,CMS:2021Yv2}.

We extend the framework to small systems, \pPb, \pO, and \OO collisions at \sqsn = 8 TeV. Generally, a stronger modification is observed for higher states ($R_{pA, AA}^{\Upsilon(1S)} > R_{pA, AA}^{\Upsilon(2S)} > R_{pA, AA}^{\Upsilon(3S)}$) even in small system. In low multiplicity events where the system size is very small, less suppression of \Ups 3 is observed due to the late formation time. In comparing different systems at the same multiplicity, a similar suppression is seen in $dN_{ch}/d\eta<25$. At higher multiplicity, the modification in \OO is weaker than that in \pPb because the energy density (temperature) in \OO is lower due to the larger system size. Regarding the elliptic flow results, we obtain very small ($<0.01$) elliptic flow for all three systems, even in high multiplicity events. It will be very interesting to compare with experimental results from the upcoming LHC run with the oxygen ion. It can provide valuable information on sources of nuclear effects on bottomonia production in small systems.

\section{Acknowledgments}

The work was supported by the National Research Foundation of Korea (NRF) grant funded by the Korea government (MSIT) under Project No. 2018R1A5A1025563 and 2020R1C1C1004985. We also acknowledge technical support from KIAF administrators at KISTI.

\clearpage

\newpage
\bibliography{main}

\end{document}